\documentclass[ALICE,manyauthors]{cernphprep}
\usepackage[comma,square,numbers,sort&compress]{natbib}
\usepackage{hyperref}
\usepackage{lineno}
\usepackage{xspace}
\usepackage[T1]{fontenc}
\begin{document}

\begin{titlepage}
\PHyear{2021}       
\PHnumber{233}      
\PHdate{7 November}  

\title{K$^0_{\rm S}$K$^0_{\rm S}$ and K$^0_{\rm S}$K$^\pm$ femtoscopy 
in pp collisions at $\sqrt{s}=$ 5.02 and 13 TeV}
\ShortTitle{K$^0_{\rm S}$K$^0_{\rm S}$ and K$^0_{\rm S}$K$^\pm$ femtoscopy}   

\Collaboration{ALICE Collaboration\thanks{See Appendix~\ref{app:collab} for the list of collaboration members}}
\ShortAuthor{ALICE Collaboration} 

\begin{abstract}
Femtoscopic correlations with the particle pair combinations K$^0_{\rm S}$K$^0_{\rm S}$
and K$^0_{\rm S}$K$^\pm$ are studied in pp collisions at $\sqrt{s}=5.02$ and 
$13$ TeV by the 
ALICE experiment. At both energies, boson source parameters are extracted for 
both pair combinations, by fitting models
based on Gaussian size distributions of the sources, to the measured two-particle correlation functions. The interaction model used for the K$^0_{\rm S}$K$^0_{\rm S}$ analysis includes quantum statistics
and strong final-state interactions through the $f_0(980)$ and $a_0(980)$ resonances. 
The model used for 
the K$^0_{\rm S}$K$^\pm$ analysis includes only the final-state interaction through the $a_0$
resonance. Source parameters extracted in the present work are compared with published
values from pp collisions at $\sqrt{s}=$ 7 TeV and the different pair combinations are found to be consistent. 
From the observation that the strength of the K$^0_{\rm S}$K$^0_{\rm S}$ correlations is
significantly greater than the strength of the K$^0_{\rm S}$K$^\pm$ correlations,
the new results are compatible with the $a_0$ resonance being a tetraquark state of the form
$(q_1,\overline{q_2}, s, \overline{s})$, where $q_1$ and $q_2$ are $u$ or $d$ quarks.

\end{abstract}
\end{titlepage}

\setcounter{page}{2} 


\section{Introduction} 
Identical boson femtoscopy, especially
identical charged $\pi^\pm\pi^\pm$ femtoscopy, has been used extensively over
the years to study experimentally the space-time geometry of the
collision region in high-energy proton and heavy-ion collisions
\cite{Lisa:2005dd}. More recently, identical-kaon femtoscopy studies have been reported
for a variety of colliding systems, energies and kaon pairs: K$^0_{\rm S}$K$^0_{\rm S}$ pairs in
Au--Au collisions at $\sqrt{s_{\rm NN}}=0.2$ TeV by the STAR collaboration~\cite{Abelev:2006gu}, 
K$^0_{\rm S}$K$^0_{\rm S}$ and K$^{\rm \pm}$K$^{\rm \pm}$ pairs in pp collisions at $\sqrt{s}=7$ TeV and
Pb--Pb collisions at $\sqrt{s_{\rm NN}}=2.76$ TeV by the ALICE collaboration~\cite{Abelev:2012ms,Abelev:2012sq,Adam:2015vja}.
Identical-kaon femtoscopy gives information about the collision region that is complementary
to that obtained in identical-pion femtoscopy in that it probes the hotter region where strange
quarks are produced and extends the momentum range over which the femtoscopy
analysis can be applied.
Also, the kaon analyses are expected to offer a cleaner signal compared to pions, as they are less affected by resonance decays. 

Non-identical kaon femtoscopy with K$^0_{\rm S}$K$^\pm$ pairs was first measured by ALICE
in pp collisions at $\sqrt{s}=7$ TeV and
Pb--Pb collisions at $\sqrt{s_{\rm NN}}=2.76$ TeV~\cite{Acharya:2018kpo,Acharya:2017jks}.
Although the space-time geometry of the kaon source can be extracted with these pairs, 
the main emphasis of
non-identical kaon femtoscopy is to obtain information about the strong final-state interaction (FSI)
between the kaons.
For the identical kaon cases the interactions are, for K$^{\rm \pm}$K$^{\rm \pm}$: quantum statistics, Coulomb interaction, and for K$^0_{\rm S}$K$^0_{\rm S}$: quantum statistics, FSI through $f_0(980)$ and $a_0(980)$ threshold resonances~\cite{Abelev:2006gu}. For the K$^0_{\rm S}$K$^{\rm \pm}$, the only FSI is through the $a_0$ resonance. Note that ``threshold resonances'', 
like the $a_0$ and $f_0$, are resonances where the sum of the masses of the decay particles is very close in value to the mass of the resonance.
A non-resonant FSI in the K$^0_{\rm S}$K$^0_{\rm S}$ pair is expected to be small compared with the resonant $f_0$ and $a_0$ FSI and can be neglected to 
first order~\cite{Abelev:2006gu,Beane:2007uh}.
The only pair-wise interaction expected for
a K$^0_{\rm S}$K$^{\rm \pm}$ pair at freeze out from the collision system is a FSI through
the $a_0$ resonance. This is because there is no quantum statistics enhancement for non-identical kaons,
no Coulomb effect since one of the kaons is neutral, and no strong FSI through the $f_0$ resonance since the
kaon pair is in an isospin-1 state, as is the $a_0$, whereas the $f_0$ is isospin-0 and thus isospin would not
be conserved.

Another feature of the K$^0_{\rm S}$K$^{\rm \pm}$ FSI through the $a_0$ resonance is that since the $a_0$ has zero strangeness, and the K$^0_{\rm S}$ is composed of a linear combination of the K$^0$ and 
${\rm \overline{K^0}}$,
only the ${\rm \overline{K^0}}$K$^+$ pair from K$^0_{\rm S}$K$^{\rm +}$ and the K$^0$K$^-$
pair from K$^0_{\rm S}$K$^{\rm -}$ can form $a_0$ resonances in order to conserve 
zero strangeness. This feature allows the
K$^0$ and ${\rm \overline{K^0}}$ sources to be studied separately. However, it was concluded in the previous ALICE K$^0_{\rm S}$K$^{\rm \pm}$ publications that there is no significant difference in the source parameters between K$^0_{\rm S}$K$^{\rm +}$ and K$^0_{\rm S}$K$^{\rm -}$ ~\cite{Acharya:2018kpo,Acharya:2017jks}.

Lastly, the K$^0_{\rm S}$K$^{\rm \pm}$ FSI allows the properties of the $a_0$ resonance itself to be studied. This is interesting in its own right since many works exist
in the literature discussing the possibility that the $a_0$ could be a 4-quark state, i.e.
a tetraquark.
It was first suggested in 1977 that experimentally-observed low-lying mesons, such as the $a_0$, are 
part of a SU(3) tetraquark
nonet using a quark model~\cite{Jaffe:1976ig}. A later follow-up calculation was published reinforcing this work using lattice QCD calculations~\cite{Alford:2000mm}. 
Since then, there have been a number of QCD studies of these mesons that fall in the categories of QCD-inspired models, for example Refs.~\cite{Santopinto:2006my,Narison:2008nj,Achasov:2017zhy,Azizi:2019kzj}, and lattice QCD calculations, for example 
Refs.~\cite{Dudek:2016cru,Briceno:2016mjc,Guo:2013nja}.
An interesting result that was found from the previous measurements in comparing the strengths, 
i.e. $\lambda$-parameters, of the K$^0_{\rm S}$K$^0_{\rm S}$ and K$^0_{\rm S}$K$^{\rm \pm}$ correlations with each other was that the strength of the K$^0_{\rm S}$K$^0_{\rm S}$ correlations is
significantly larger than the strength of the K$^0_{\rm S}$K$^{\rm \pm}$ correlations
measured in $\sqrt{s}=7$ TeV pp collisions. It was
suggested that this could be an indication that the $a_0$ is a tetraquark 
state~\cite{Acharya:2018kpo,Acharya:2017jks}.

In light of the interesting results from the $\sqrt{s}=7$ TeV pp measurements,  the main motivations to extend the measurements to $\sqrt{s}=$ 5.02 and 13 TeV 
pp collisions are the following:
\begin{itemize}
\item In these new measurements, we investigate the collision-energy dependence of the $\lambda$ difference between K$^0_{\rm S}$K$^0_{\rm S}$ and 
K$^0_{\rm S}$K$^{\rm \pm}$. A lack of a dependence on the center-of-mass energy would be
consistent with the tetraquark interpretation of the $a_0(980)$.
 
\item The previous paper compared results for $\lambda$ that were obtained and published seven years apart, i.e. K$^0_{\rm S}$K$^0_{\rm S}$ in PLB from 2012~\cite{Abelev:2012ms} and K$^0_{\rm S}$K$^{\rm \pm}$ in PLB from 2019~\cite{Acharya:2018kpo}, and that were analyzed in different ways. The K$^0_{\rm S}$K$^0_{\rm S}$ analysis from the 2012 paper was done in several multiplicity ranges which had to be averaged in order to compare with the minimum-bias 2019 K$^0_{\rm S}$K$^{\rm \pm}$ result. However, in the present paper the new K$^0_{\rm S}$K$^0_{\rm S}$ and K$^0_{\rm S}$K$^{\rm \pm}$ analyses were done at the same time and using the same kinematic ranges. 
By carrying out simultaneous measurements of K$^0_{\rm S}$K$^0_{\rm S}$ and K$^0_{\rm S}$K$^{\rm \pm}$ this results in a better comparison with each other.

\item In this new analysis a detailed calculation of the effect of long-lived resonances on the $\lambda$ parameter is presented to better establish that this contamination is not responsible for the $\lambda$ difference. 
\end{itemize}

\section{Description of experiment and data selection}
Data taken by the ALICE experiment~\cite{Aamodt:2008zz} in the LHC Run 2
period (2015--2018) were employed in the present analysis. This analysis used both 
$\sqrt{s}=$ 5.02 TeV and 13 TeV reconstructed minimum bias triggered pp collisions,
giving about $0.5\times10^9$ and $1.5\times10^9$ events, respectively.
Monte Carlo (MC) simulations were used for
determining selection values, momentum resolution and purity studies, and for the baseline
underlying the signal for the case of the K$^0_{\rm S}$K$^0_{\rm S}$ analyses.
In the MC calculations, particles from pp collision events simulated by the general-purpose generator  PYTHIA8~\cite{Sjostrand:2006za} with the Monash 2013 tune~\cite{Skands:2014pea}
were transported through a GEANT3~\cite{Brun:1994aa} model of the ALICE detector.
The total numbers of MC events used in the $\sqrt{s}=$ 5.02 and 13 TeV analyses
were about
$0.7\times 10^9$ and $1.2\times10^9$, respectively.

The V0 detectors, which consist of two arrays of scintillators located along the beamline and 
covering the full
azimuth~\cite{Abelev:2013vea,Abelev:2013qoq} were used for triggering and event selection.
Charged particles were reconstructed and identified with the central barrel detectors located 
within a solenoid magnet with a field strength of magnitude $B=0.5$ T.
Charged particle tracking was performed using the 
Time Projection Chamber (TPC)~\cite{Alme:2010ke} 
and the Inner Tracking System (ITS)~\cite{Aamodt:2008zz}. 
The momentum determination for K$^\pm$ was made using only the TPC. The ITS allowed for high spatial resolution in determining the primary collision vertex, which was used to constrain the TPC tracks.
An average momentum resolution of less than 10 MeV/$c$ was typically obtained for the charged tracks
of interest in this analysis~\cite{Alessandro:2006yt}.
The primary vertex was obtained
from the ITS, the position being constrained along the beam direction to be within $\pm10$ cm of the center of the ALICE detector.
In addition to the standard track quality selections~\cite{Alessandro:2006yt}, the selections based on the quality of track fitting and the number of detected tracking points in the TPC were used to ensure that only well-reconstructed tracks were taken into account in the analysis~\cite{Abelev:2014ffa,Alme:2010ke,Alessandro:2006yt}.

Particle Identification (PID) for reconstructed tracks was carried out using both the TPC and the Time-Of-Flight (TOF) detectors in the pseudorapidity range $|\eta| < 0.8$~\cite{Abelev:2014ffa,Akindinov:2013tea}.
For the PID signal from both detectors, a value ($N_{\sigma}$) was assigned to each track denoting the number of standard deviations between the measured track information and expected values, assuming a mass hypothesis, divided by the detector resolution
\cite{Adam:2015vja,Abelev:2014ffa,Akindinov:2013tea,Alessandro:2006yt}.
For TPC PID, a parametrized Bethe-Bloch formula was used to calculate the specific energy 
loss $\left<{\rm d}E/{\rm d}x\right>$ in the detector expected for a particle with a given charge, mass and momentum. For PID with TOF, the particle mass was used to calculate the expected time-of-flight as a function of track length and momentum. 

Other event selection criteria were also applied.
The event must have one accepted possible K$^0_{\rm S}$K$^0_{\rm S}$ or K$^0_{\rm S}$K$^{\rm \pm}$ pair.
Pile-up events were rejected using the standard ALICE pile-up rejection
method~\cite{Abelev:2014ffa}.
Pile-up effects were also investigated by performing the analysis using only low-luminosity data-taking
periods. No significant difference was found in the extracted $R$ and $\lambda$ parameters compared with the higher count-rate runs used.

\subsection{Kaon selection}
The methods used to select and identify individual K$^0_{\rm S}$ and K$^{\rm \pm}$ particles are the same as those used for the ALICE K$^0_{\rm S}$K$^0_{\rm S}$~\cite{Abelev:2012ms}
and K$^{\rm \pm}$K$^{\rm \pm}$~\cite{Abelev:2012sq} analyses in pp collisions at 
$\sqrt{s}=7$ TeV, and are described in the following sections.

\subsubsection{K$^0_{\rm S}$ reconstruction}
Using an invariant mass technique, the neutral K$^0_{\rm S}$ vertices and parameters are reconstructed and calculated from pairs of detected $\pi^+$ $\pi^-$ tracks. Single-particle selection criteria for the K$^0_{\rm S}$
and the pions, for example particle momentum ($p$), transverse momentum ($p_T$), 
and pseudorapidity ($\eta$), 
are shown in Table~\ref{tab:singleKcuts}.

\begin{table}
 \centering
 \caption{Single-particle selection criteria.}
 \begin{tabular}{| l | c |}
  \hline
  {\bf Neutral kaon selection} & {\bf Value} \\ \hline
  Daughter $p_{\rm T}$ & $> 0.15$ GeV/$c$ \\  
  Daughter $|\eta|$ &  $< 0.8$ \\
  Daughter DCA (3D) to primary vertex & $>0.4$ cm \\
  Daughter TPC $N_{\rm \sigma}$ &  $< 3$ \\
  Daughter TOF $N_{\rm \sigma}$ (for $p > 0.8$ GeV/$c$) &  $< 3$ \\ \hline
  $\left|\eta\right|$ &  $< 0.8$ \\
  DCA (3D) $\pi^+$ to $\pi^-$ & $< 0.3$ cm \\
  DCA (3D) of K$^0_{\rm S}$ to primary vertex & $< 0.3$ cm \\
  Decay length (3D, lab frame) & $< 30$ cm \\
  Decay radius (2D, lab frame) & $> 0.2$ cm \\
  Cosine of pointing angle & $> 0.99$ \\
  Invariant mass & $0.485 < m < 0.510$ GeV/$c^{\rm 2}$ \\ \hline
  {\bf Charged kaon selection} & {\bf Value} \\ \hline
  $p_{\rm T}$ & $0.15 < p_{\rm T} < 1.2$ GeV/$c$ \\
  $|\eta|$ &  $< 0.8$ \\
  Transverse DCA to primary vertex & $<2.4$ cm \\
  Longitudinal DCA to primary vertex & $<3.0$ cm \\
  $N_{\rm \sigma}^{TOF}$ with valid TOF signal and $p>0.5$ GeV/$c$ & $<2$  \\
  $N_{\rm \sigma}^{TPC}$ if no TOF signal for all $p_{\rm T}$ & $< 2$  \\
  Kalman fit $\chi^2/N^{\rm clus}$ & $\leq 4$ \\
  \hline
  \end{tabular}
 
  \label{tab:singleKcuts}
\end{table}

\noindent Most of the topological selection criteria ($\pi^+$ $\pi^-$ distance-of-closest-approach (DCA), 
$\pi$-vertex DCA, K$^0_{\rm S}$ DCA, and decay length) were chosen to optimize purity and 
statistical significance.  If two reconstructed K$^0_{\rm S}$ particles share a daughter track, both are removed from the analysis. The selection criteria in this analysis are comparable to or stricter than those in other K$^0_{\rm S}$ analyses; strict selection criteria are favored to increase the sample purity.

A candidate K$^0_{\rm S}$ vertex with a reconstructed invariant mass within 
$0.485 < m(\pi^+\pi^-) < 0.510$ GeV/$c^2$ is identified as a K$^0_{\rm S}$. In this range,
the single-K$^0_{\rm S}$ purity is measured to be $98\pm1\%$ for 
the $k_{\rm T}$ interval of $0.5<k_T<0.7$ GeV/$c$, 
where $k_{\rm T}=|\overrightarrow{p_{\rm T1}}+\overrightarrow{p_{\rm T2}}|/2$, and 
where $\overrightarrow{p_{\rm T1}}$ and
$\overrightarrow{p_{\rm T2}}$ are the transverse momenta of the particles in the pair.
The purity here is defined as Signal/(Signal + Background) and is calculated by fitting a fourth-order polynomial 
to the background in the combined invariant mass intervals 0.4--0.45 GeV/$c^2$ and 
0.55--0.6 GeV/$c^2$
and using the bin contents of the invariant mass histogram as the ``Signal + Background''. 
No selection on $p_{\rm T}$ is employed in this analysis for K$^0_{\rm S}$. Having a pair purity less than unity will be reflected in the lowering of the 
$\lambda$ parameter, which can later be corrected for purity, however the K$^0_{\rm S}$
purity is very close to unity for this analysis.

\subsubsection{K$^{\rm \pm}$ identification}
As mentioned above,
charged kaons are selected using the TPC and TOF detectors with the same methods as
employed in Refs.~\cite{Abelev:2012sq,Adam:2015vja}.  
The quality of the track is 
determined by the $\chi^2/N^{\rm clus}$ value for the Kalman fit to the reconstructed position of the TPC clusters ($N^{\rm clus}$ is the number of clusters attached to the track). The track is rejected if the value is larger than 4.0.
The selection criteria used for the charged kaon selection in the TPC and TOF are shown in Table~\ref{tab:singleKcuts}. In the table, $N_{\rm \sigma}^{TPC}$ and $N_{\rm \sigma}^{TOF}$
are the numbers of standard deviations the TPC energy-loss and TOF signal are away from their predicted values divided by detector resolution, respectively.

The average charged kaon purity is found using PYTHIA8 MC simulations to be $91\pm1\%$ in the $k_{\rm T}$ range used in this 
analysis, i.e. $0.5<k_{\rm T}<0.7$ GeV/$c$. This is
in agreement with the charged kaon purity found by the ALICE collaboration in Ref.~\cite{Adam:2015vja}.

\subsection{Two-track selection}
Experimental two-track effects, such as the merging of two real tracks into one reconstructed track and the splitting of one real track into two reconstructed tracks, is a challenge for femtoscopic studies.
These effects are observed for tracks with small average separation in the TPC. 
For each pair of like-sign tracks, which could be pions from two K$^0_{\rm S}$ decays, or the pion from a 
K$^0_{\rm S}$ decay and the same-charged K$^{\rm \pm}$ track, 
the distance between the 
tracks was calculated at up to nine positions throughout the TPC (every 20 cm along the radial direction from 85 cm to 245 cm) and then averaged.
When comparing the distribution of the average separation of track pairs from single events with the distribution from pairs constructed of tracks from different events (mixed events), a splitting enhancement is seen in the same-events for average separations approaching zero.
For the distribution of mixed-event tracks, the primary vertex position for each track was subtracted from each track point to mock them up as coming from the same event.
To minimize this splitting effect, this analysis demanded that the tracks must have an average TPC separation of at least 13 cm.

\section{Two-particle correlation function}
This analysis studies the momentum correlations of K$^0_{\rm S}$K$^0_{\rm S}$ and
K$^0_{\rm S}$K$^{\rm \pm}$  pairs using the two-particle correlation function, defined as $C(k^*)=A(k^*)/B(k^*)$, where $A(k^*)$ is the measured distribution of real pairs from the same event and $B(k^*)$ is the reference distribution of pairs from mixed events. The quantity $k^*$ is the momentum of one of the particles in the pair rest frame, and for the general case of two
particles with unequal mass, $m_1$ and $m_2$, is given by

\begin{equation}
k^*=\sqrt{\frac{w^2-m_1^2m_2^2}{2w+m_1^2+m_2^2}}
\label{kstar}
\end{equation}
where,
\begin{equation}
w\equiv (q_{inv}^2+m_1^2+m_2^2)/2.
\label{kstar2}
\end{equation}

The square of the invariant momentum difference 
$q_{\rm inv}^2=|\vec{p_1}-\vec{p_2}|^2-|E_1-E_2|^2$ is most conveniently evaluated with the momenta and energies of the two particles measured in the lab frame. Note that $m_1=m_2$ gives 
$k^*=q_{\rm inv}/2$.
The denominator $B(k^*)$ is formed by mixing particles from each event with particles from ten other events in the same $z$-vertex bin (2 cm width) and of similar event multiplicity.  
A $k^*$ bin size of 
20 MeV/$c$ was used in all cases.

As mentioned earlier, correlation functions are calculated for minimum bias events and a
$k_{\rm T}$ range from 0.5--0.7 GeV/$c$. This closely reproduces the conditions for the
kaon femtoscopy measurements with K$^0_{\rm S}$K$^0_{\rm S}$ and K$^0_{\rm S}$K$^\pm$ pairs published by ALICE
for pp collisions at $\sqrt{s}=7$ TeV with which the present results will be compared~\cite{Abelev:2012ms,Acharya:2018kpo}. The $k_{\rm T}$ range used
encompasses the peak in the $k_{\rm T}$ distributions at each collision energy. 
Also, the pseudorapidity density
of charged particles at midrapidity, $dN_{ch}/d\eta$, is found to be small in pp collisions and
has a weak dependence on $\sqrt{s}$, measured to be $5.91\pm0.45$, $6.01^{+0.20}_{-0.12}$, and
$7.60\pm0.50$ for $\sqrt{s}=5.02$, 7 and 13 TeV, 
respectively, where the uncertainties are the statistical and systematic uncertainties added in quadrature~\cite{Acharya:2019idg,Acharya:2019mzb,Aamodt:2010pp}.

\begin{figure}[]
\centering
\includegraphics[width=75mm]{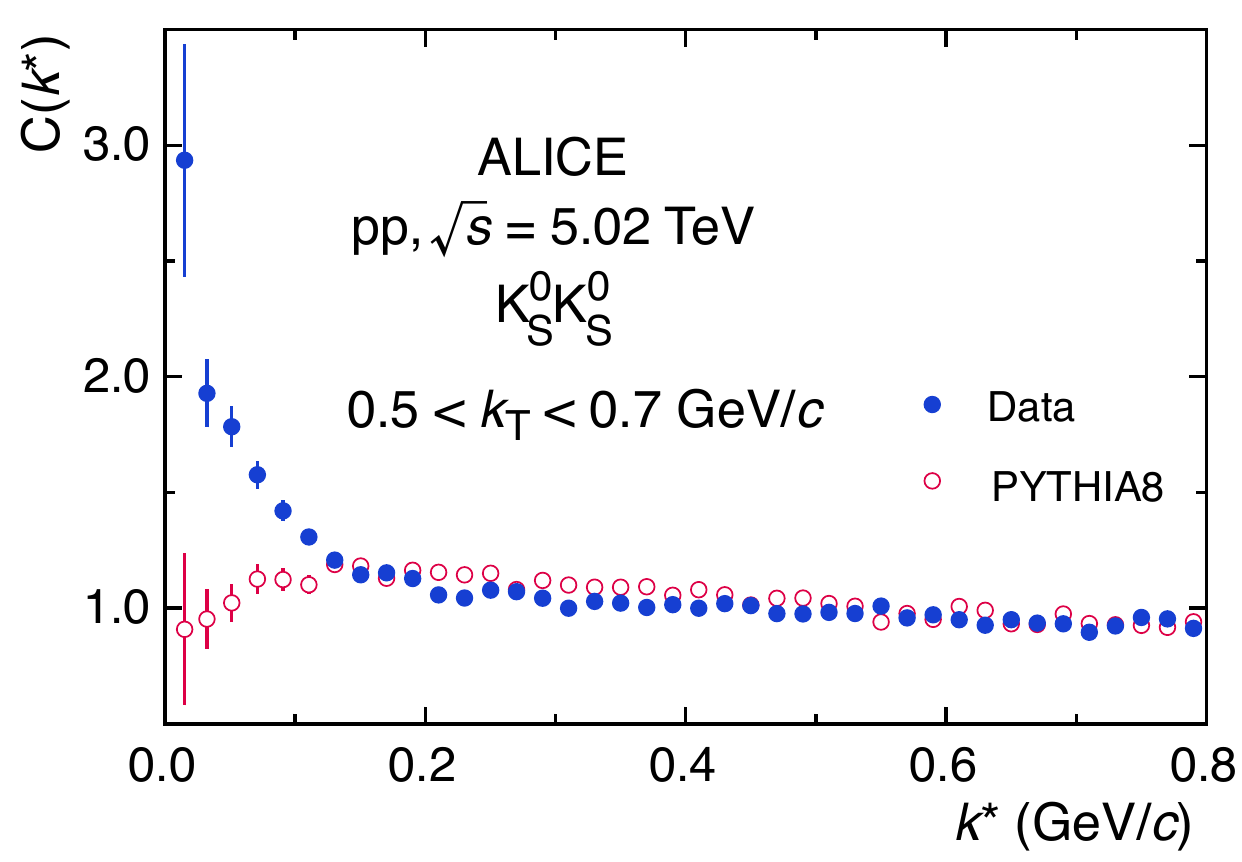}
\hspace{1mm}
\includegraphics[width=75mm]{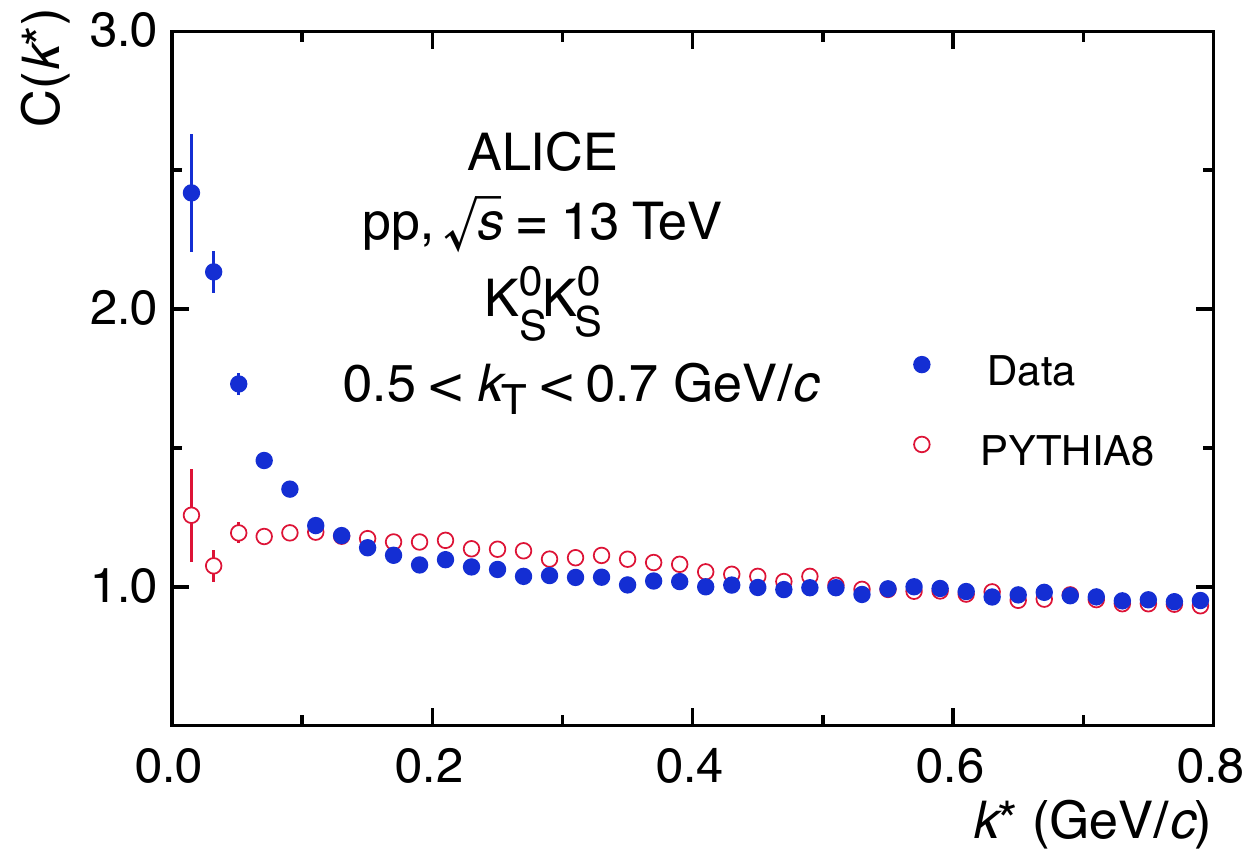}
\caption{Example K$^0_{\rm S}$K$^0_{\rm S}$ correlation functions 
along with the resulting distributions from PYTHIA8 simulations
in pp collisions at $\sqrt{s}=5.02$ (left) and $13$ TeV (right).}
\label{fig:rawKKCF5a13MC}
\end{figure}

\begin{figure}[]
\centering
\includegraphics[width=75mm]{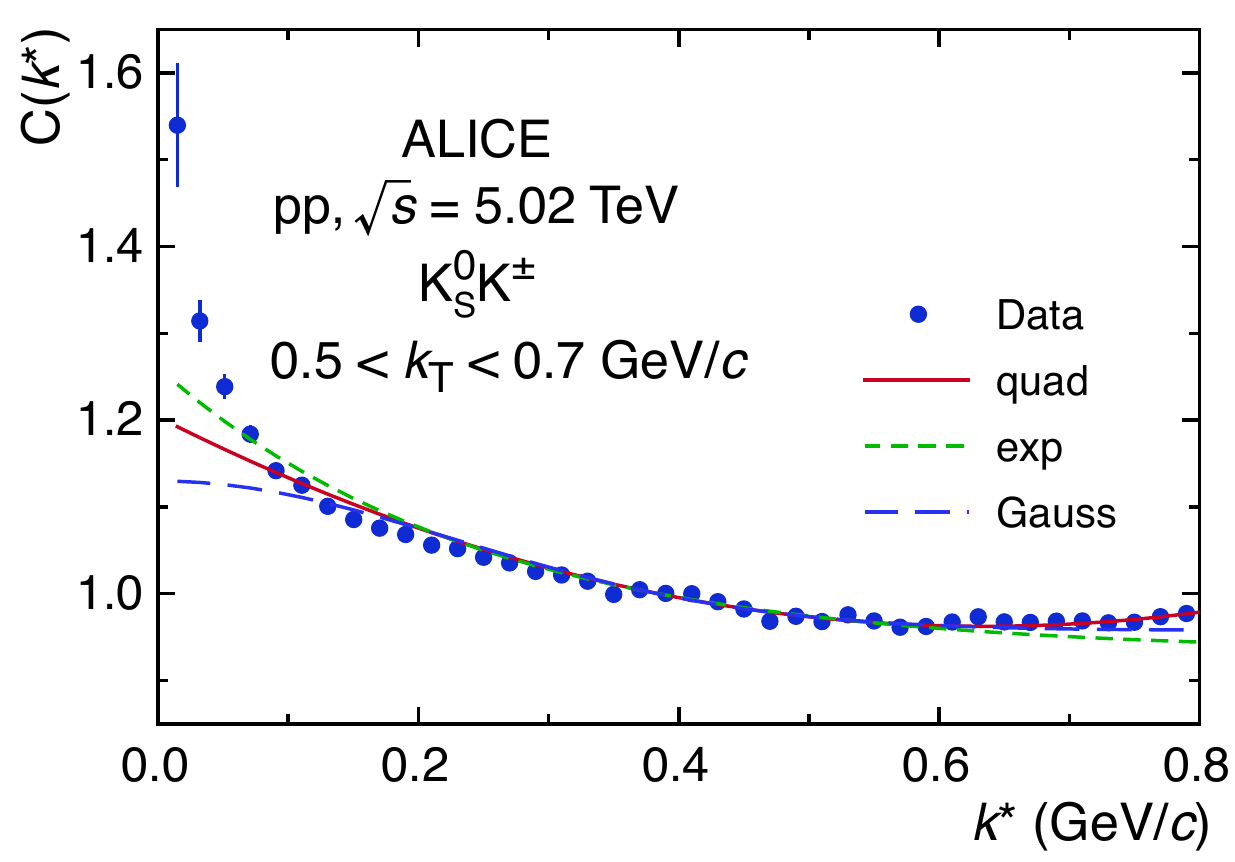}
\hspace{1mm}
\includegraphics[width=75mm]{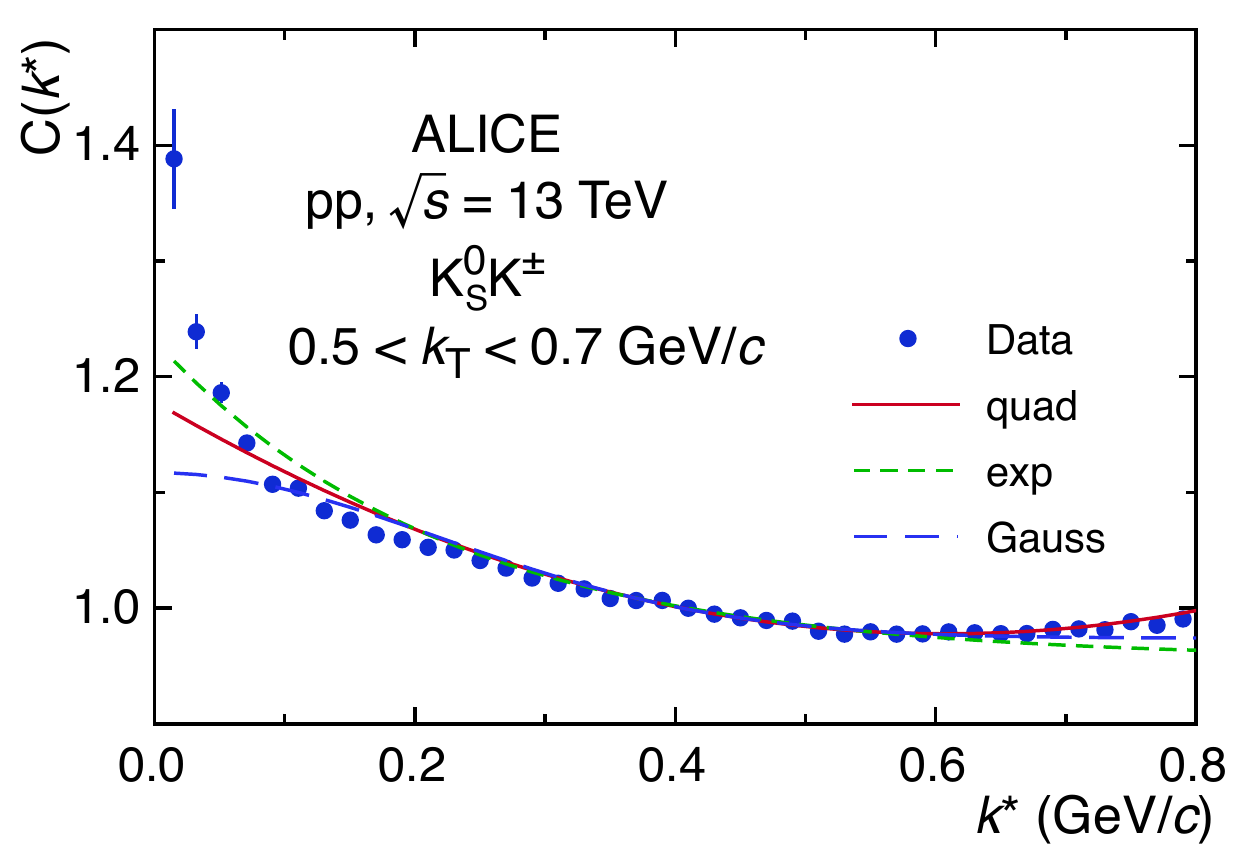}
\caption{Example K$^0_{\rm S}$K$^\pm$ correlation functions plotted with fits of
	Eqs.~\ref{quad}, ~\ref{exp} and ~\ref{gauss} in pp collisions at
	$\sqrt{s}=5.02$ (left) and $13$ TeV (right).}
\label{fig:rawKKchCF5a13}
\end{figure}

Figure~\ref{fig:rawKKCF5a13MC} shows an example of
raw experimental K$^0_{\rm S}$K$^0_{\rm S}$ correlation functions
along with the resulting distributions from PYTHIA8 simulations
normalized in the $k^*$ region 0.6--0.8 GeV/$c$ for $\sqrt{s}=$ 5.02 and 13 TeV. 
Note that the PYTHIA8 calculations do not contain FSI or femtoscopic correlations.
Figure~\ref{fig:rawKKchCF5a13}
shows an example of raw experimental K$^0_{\rm S}$K$^\pm$ correlation functions plotted with
baseline fits for various functions (see below) for $\sqrt{s}=$ 5.02 and 13 TeV. 
The raw correlation functions from the data are enhanced
for $k^*<0.1$ GeV/$c$ due to quantum statistics and the FSI of the $f_0$ and $a_0$ and slightly
suppressed in the region $0.1<k^*<0.4$ GeV/$c$ due to the FSI for
K$^0_{\rm S}$K$^0_{\rm S}$.
For K$^0_{\rm S}$K$^{\rm \pm}$ the FSI of the $a_0$ produces similar but smaller enhancements 
and suppressions in the same general $k^*$ regions.
For $k^*>0.4$ GeV/$c$ a non-flat baseline is observed in both cases. 
PYTHIA8 fairly describes the non-flat baseline of
the experimental correlation functions for K$^0_{\rm S}$K$^0_{\rm S}$, and is thus used to take out the effect of the non-flat baseline by dividing the raw experimental correlation functions by
the PYTHIA8 correlation functions in those cases. This is a similar method as was used for the
$\sqrt{s}=$ 7 TeV pp K$^0_{\rm S}$K$^0_{\rm S}$ measurements~\cite{Abelev:2012ms}.

\begin{figure}[]
\centering
\includegraphics[width=75mm]{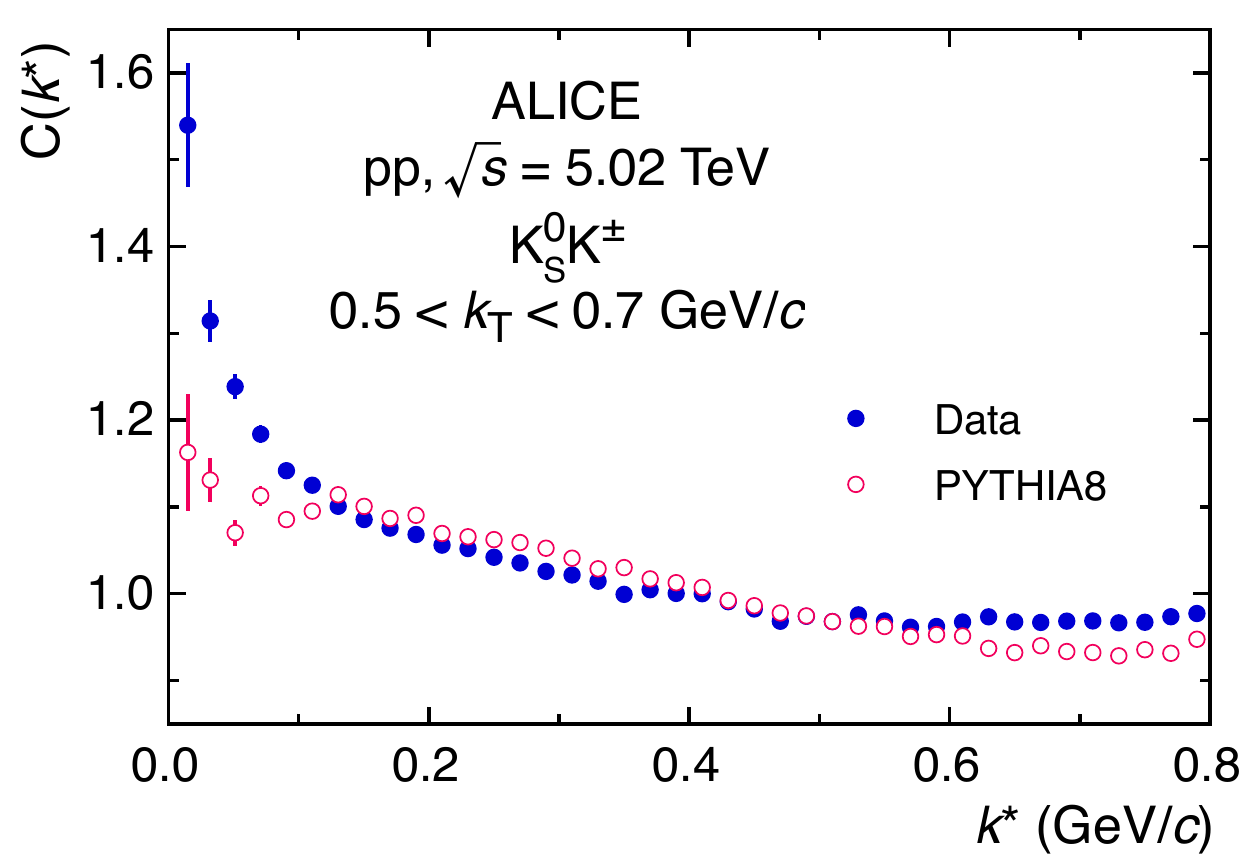}
\hspace{1mm}
\includegraphics[width=75mm]{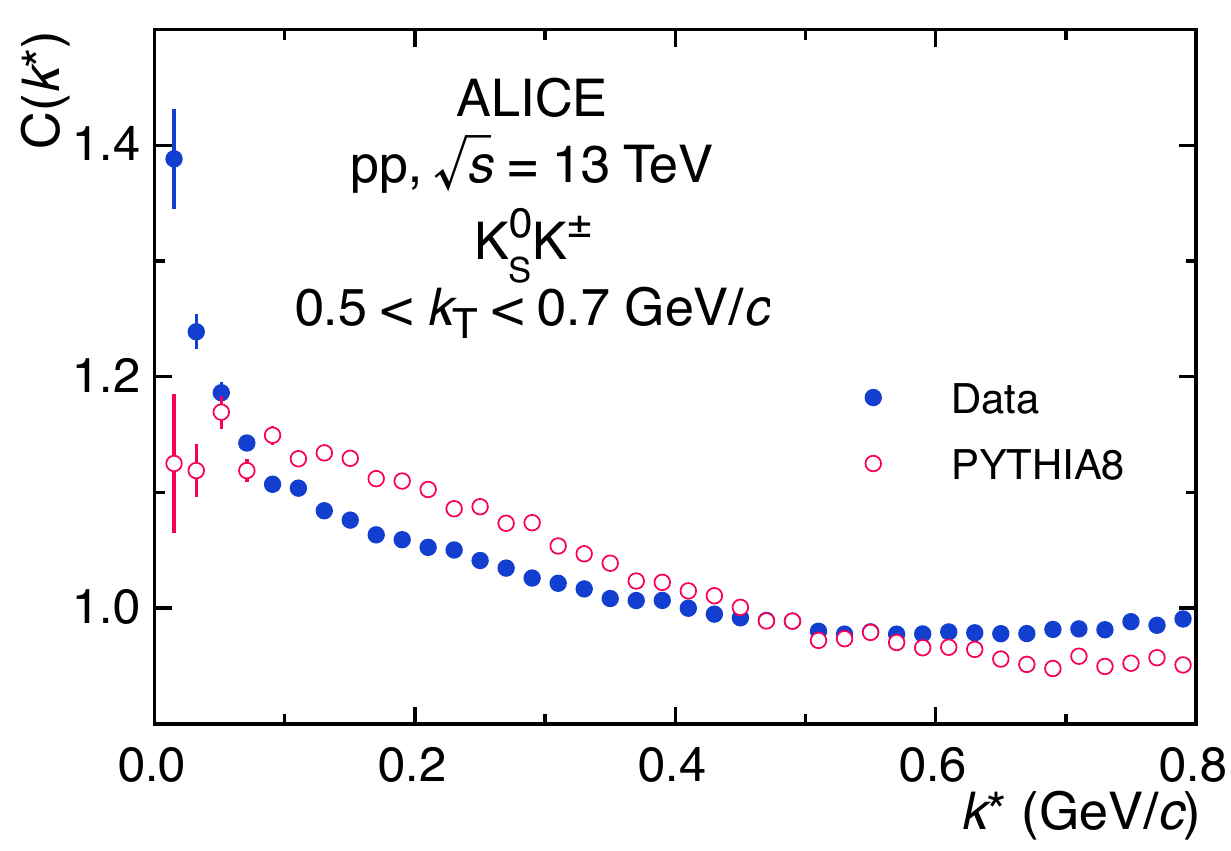}
\caption{Example K$^0_{\rm S}$K$^\pm$ correlation functions
along with the resulting distributions from PYTHIA8 simulations
in pp collisions at $\sqrt{s}=5.02$ (left) and 13 TeV (right).}
\label{fig:K0Kch13MC}
\end{figure}

Unlike the case for K$^0_{\rm S}$K$^0_{\rm S}$, which has a relatively large signal compared with
the baseline, 
for the K$^0_{\rm S}$K$^{\rm \pm}$ correlation functions PYTHIA8 is not sensitive enough to model
the baseline sufficiently well with respect to the significantly smaller enhancement and
suppression produced by
the FSI of the $a_0$ alone. 
Examples of this are shown in Fig.~\ref{fig:K0Kch13MC}, which compares the raw 
$\sqrt{s}=5.02$ and 13 TeV
pp correlation functions from data with those from PYTHIA8. 
For these correlation functions, quadratic, exponential and Gaussian
functions are used to model the baseline, as was done in Ref.~\cite{Acharya:2018kpo} for the 
measurement in pp collisions at $\sqrt{s}=$ 7 TeV, of the forms

\begin{equation}
C_{\rm quadratic}(k^*) = a(1-bk^*+ck^{*2})
\label{quad}
\end{equation}

\begin{equation}
C_{\rm exponential}(k^*) = a(1+b\exp(-ck^*))
\label{exp}
\end{equation}

\begin{equation}
C_{\rm Gaussian}(k^*) = a(1+b\exp(-ck^{*2}))
\label{gauss}
\end{equation}

where $a$, $b$ and $c$ are parameters that are fitted to the
experimental $C(k^*)$ simultaneously with the FSI 
model (see Section 4). 
As shown in Fig.~\ref{fig:rawKKchCF5a13}, the quadratic, Gaussian and exponential functions all describe the data well in the
$k^*$ range of $\sim$0.3--0.6~GeV/$c$.

\section{Fitting the correlation functions to extract the source parameters}

\subsection{K$^0_{\rm S}$K$^0_{\rm S}$}
The K$^0_{\rm S}$K$^0_{\rm S}$ correlation functions were fitted with the Lednick\'{y}
parameterization~\cite{Abelev:2006gu} which incorporates quantum statistics with strong FSI.
FSI arise in the K$^0_{\rm S}$K$^0_{\rm S}$ channels due to the near-threshold resonances, 
$a_0(980)$ and $f_0(980)$. This parameterization is based on the model by R. Lednick\'{y} and V.L. Lyuboshitz~\cite{Lednicky:1981su,Lednicky:2005af}.

The general form of the fit function is:

\begin{equation}
C_{\rm Lednicky}(k^*)=1 + \lambda e^{-4k^{*2}R^2} + \lambda\alpha\left[\left|\frac{f(k^*)}{R}\right|^2+\frac{4\mathcal{R}f(k^*)}{\sqrt{\pi}R}F_1(2k^* R)-\frac{2\mathcal{I}f(k^*)}{R}F_2(2k^* R) + \Delta C\right]
\label{eq:fit1}
\end{equation}
where
\begin{equation}
F_1(z)=\int^{z}_{0}dx\frac{e^{x^2-z^2}}{z}; \hspace{3mm} F_2(z)=\frac{1-e^{-z^2}}{z}.
\label{eq:fit2}
\end{equation}
The scattering amplitude is
\begin{equation}
f(k^*)=\frac{f_0(k^*)+f_1(k^*)}{2}
\label{eq:fit3}
\end{equation}
where
\begin{equation}
f_{\rm I}(k^*) = \frac{\gamma_{\rm I}}{m_{\rm I}^2-s-i(\gamma_{\rm I} k^*+\gamma_{\rm I}'k_{\rm I}')},
\label{eq:fit4}
\end{equation}

$f(k^*)$ is the s-wave K$^0\overline{\rm K^0}$ scattering amplitude whose contributions are the isoscalar $f_0$ and isovector $a_0$ resonances; $\alpha$ is set to 0.5 assuming symmetry in K$^0$ and 
$\overline{\rm K^0}$ production; $R$ is the radius parameter; and $\lambda$ is the correlation strength. In Eq.~\ref{eq:fit4}, $I=$0 or 1 for the $f_0$ or $a_0$, $m_{\rm I}$ is the mass of the resonance, and 
$\gamma_{\rm I}$ and $\gamma_{\rm I}'$ are the couplings of the resonances to their decay channels. Also, $s=4(m_K^2+k^{*2})$ and $k_{\rm I}'$ denotes the momentum in the second decay channel. 
The K$^0\overline{\rm K^0}$ s-wave scattering amplitude depends on the $f_0$ and $a_0$ resonance mass and decay couplings, which have been measured~\cite{Achasov:2002ir}. The parameter set used in the present analysis is shown in Table~\ref{tab:respar}.
The quantity $\Delta C$ is a correction for small source sizes found in pp collisions~\cite{Abelev:2006gu}, and is given by:
\begin{equation}
\Delta C = \frac{1}{\sqrt{\pi}R^3}\left[ |f_0(k^*)|^2(\frac{3}{\gamma_0}+\frac{1}{\gamma_1})+|f_1(k^*)|^2(\frac{1}{\gamma_0}+\frac{3}{\gamma_1})\right].
\label{eq:delc}
\end{equation}

\begin{table}
 \centering
  \caption{The $f_0$ and $a_0$ masses and coupling parameters used in the present
  analysis, all in GeV.}
 \begin{tabular}{| c | c | c | c | c | c |}
  \hline
  $m_{f_0}$ & $\gamma_{f_0K\bar{K}}$ & $\gamma_{f_0\pi\pi}$ & $m_{a_0}$ & $\gamma_{a_0K\bar{K}}$ & $\gamma_{a_0\pi\eta}$ \\ \hline
  0.967 & 0.34 & 0.089 & 1.003 & 0.8365 & 0.4580 \\  
  \hline
  \end{tabular}
  \label{tab:respar}
\end{table}

\begin{figure}[]
\centering
\includegraphics[width=75mm]{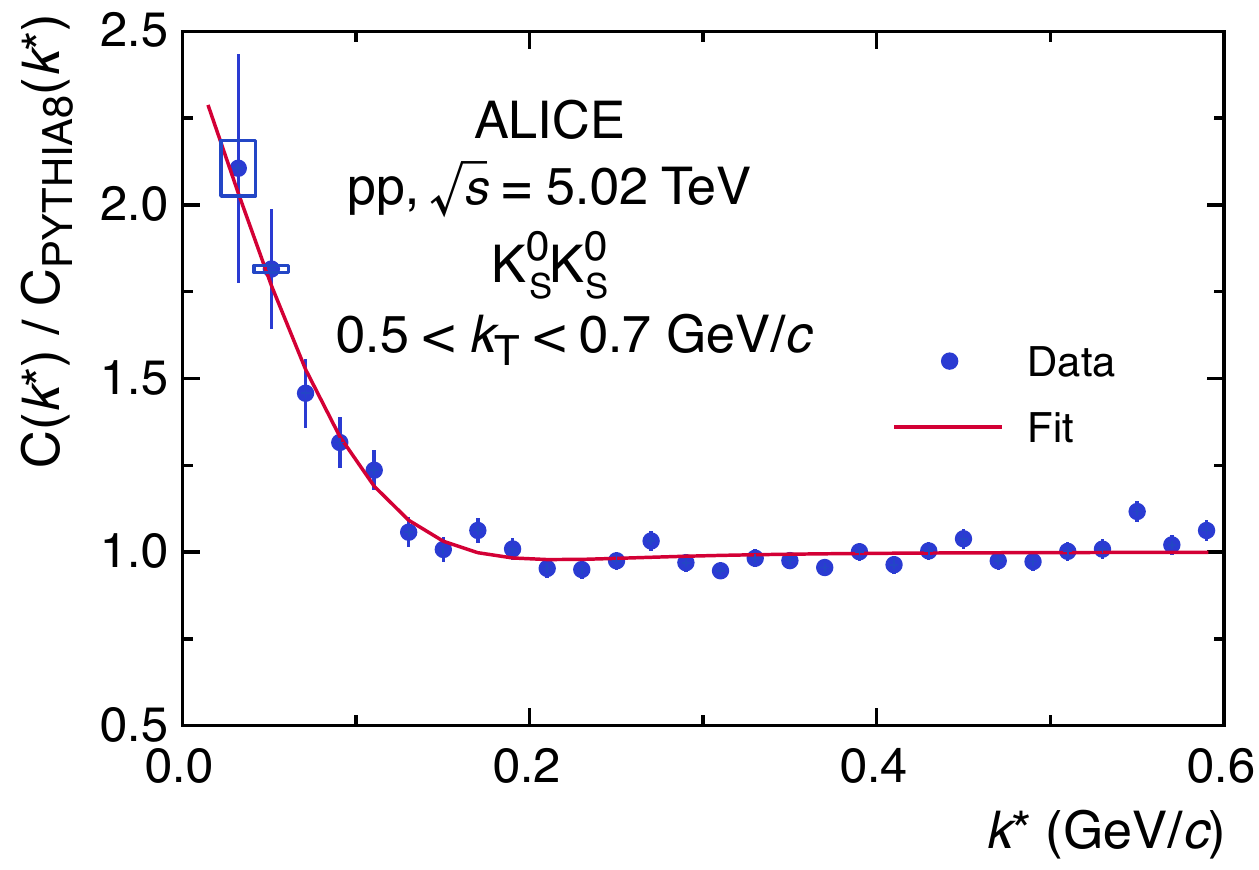}
\hspace{1mm}
\includegraphics[width=75mm]{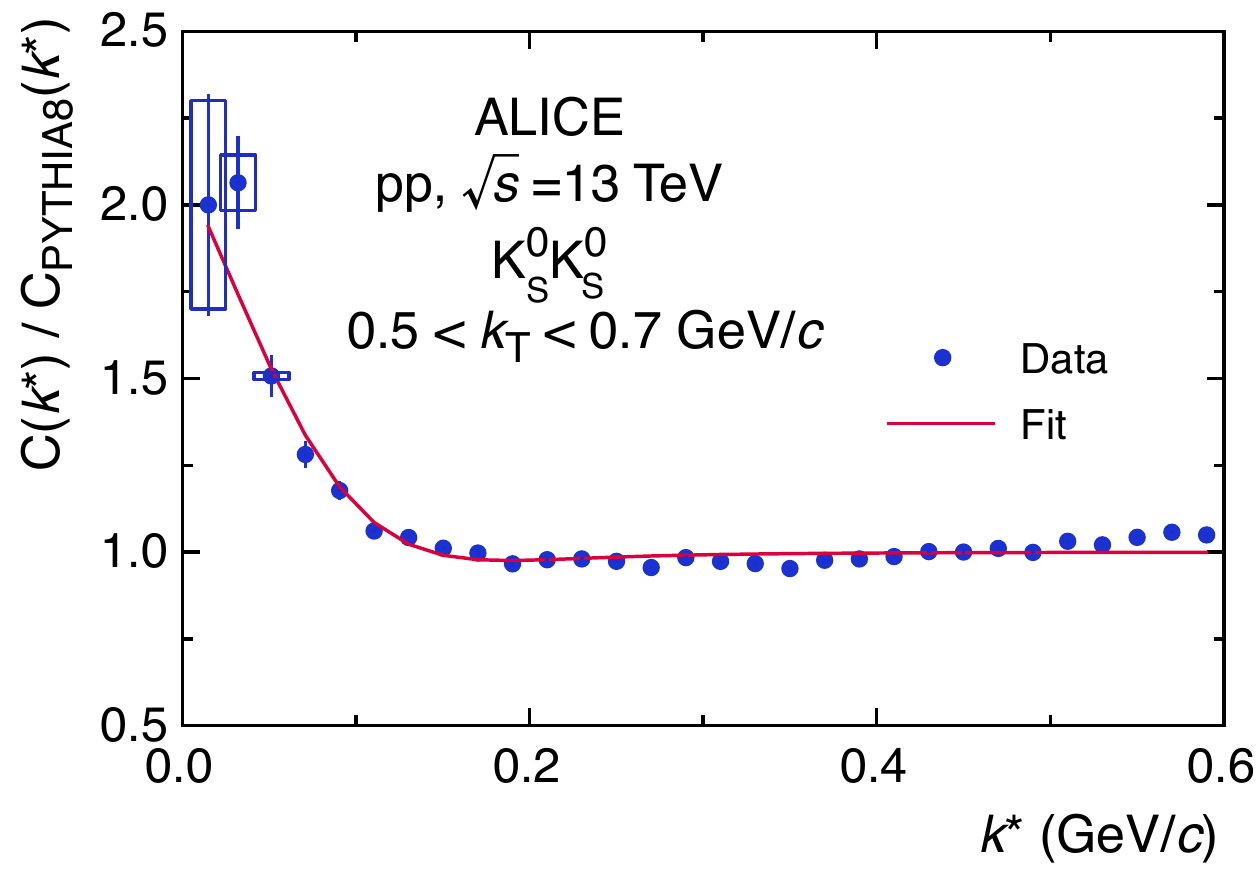}
\caption{Example fits of Eq.~\ref{eq:fit1} to the ratio of the data to
PYTHIA8 correlation functions for K$^0_{\rm S}$K$^0_{\rm S}$ in pp collisions at $\sqrt{s}=$ 
5.02 TeV (left) and 13 TeV (right). Statistical uncertainties are shown as error bars and systematic uncertainties are shown as boxes.}
\label{fig:KKCFfit1}
\end{figure}

Figure~\ref{fig:KKCFfit1} shows example fits of Eq.~\ref{eq:fit1} 
to the ratio of the data to
PYTHIA8 correlation functions for K$^0_{\rm S}$K$^0_{\rm S}$ measured in pp collisions at
$\sqrt{s}=$ 5.02 TeV and 13 TeV.
Statistical uncertainties are shown as error bars, and systematic uncertainties are shown as boxes. 
The statistical uncertainties from PYTHIA8 were propagated to those on the data points.
The fits of Eq.~\ref{eq:fit1} to the correlation function ratios provide a good description of the data,
typically giving $\chi^2/$ndf values close to unity.
The $\chi^2/$ndf values of the fits to the K$^0_{\rm S}$K$^0_{\rm S}$
correlation functions are 1.3 and 2.5, respectively, for the left and the right figures. The large
$\chi^2/$ndf value for the fit shown in the right figure mostly reflects a combination of the small
statistical uncertainties in the data and the deviation of the fit in the region $k^*>0.5$ GeV/c.

\subsection{K$^0_{\rm S}$K$^{\rm \pm}$}
\label{subsec:fitting}
The K$^0_{\rm S}$K$^{\rm \pm}$ correlation functions were fitted with the expression:

\begin{equation}
C(k^*) = C_{\rm Lednicky2}(k^*)C_{\rm baseline}(k^*)
\label{eq:corrfit}
\end{equation}

where $C_{\rm Lednicky2}(k^*)$ is a modified version of Eq.~\ref{eq:fit1}, and $C_{\rm baseline}(k^*)$ is Eq.~\ref{quad}, Eq.~\ref{exp} or Eq.~\ref{gauss}.

The modified form of the Lednick\'{y} FSI fit function used is:

\begin{equation}
C_{\rm Lednicky2}(k^*)=1+\left( \frac{\lambda\alpha}{2}\right)\left[\left|\frac{f(k^*)}{R}\right|^2+\frac{4\mathcal{R}f(k^*)}{\sqrt{\pi}R}F_1(2k^* R)-\frac{2\mathcal{I}f(k^*)}{R}F_2(2k^* R)+\Delta C'\right].
\label{eq:fit2a}
\end{equation}
The scattering amplitude is:
\begin{equation}
f(k^*) = \frac{\gamma_{a_0\rightarrow K\overline{K}}}{m_{a_0}^2-s-i(\gamma_{a_0\rightarrow K\overline{K}} k^*+\gamma_{a_0\rightarrow \pi\eta}k_{\pi\eta})}.
\label{eq:fit4a}
\end{equation}

Note that the form of the FSI term in
Eq.~\ref{eq:fit2a} differs from the form of the FSI term for K$^0_{\rm S}$K$^0_{\rm S}$ correlations, Eq.~\ref{eq:fit1},  by a factor of $1/2$ due to the non-identical particles in K$^0_{\rm S}$K$^{\rm \pm}$ correlations and thus the absence of the requirement to symmetrize the wavefunction.
The K$^0$K$^-$ or $\overline{\rm K^0}$K$^+$ s-wave scattering amplitude depends only on the $a_0$ resonance mass and decay couplings. The ones used in this analysis are
shown in Table~\ref{tab:respar}. The correction due to small source sizes, $\Delta C'$, now becomes:
\begin{equation}
\Delta C' = \frac{2}{\sqrt{\pi}R^3}\frac{|f(k^*)|^2}{\gamma_{a_0\rightarrow K\overline{K}}}.
\label{eq:delcp}
\end{equation}

The fitting strategy is to make a 5-parameter fit of Eq.~\ref{eq:corrfit} to the 
K$^0_{\rm S}$K$^{\rm \pm}$ experimental correlation
functions to extract $R$, $\lambda$, $a$, $b$ and $c$ for each
baseline functional form.

\begin{figure}[]
\centering
\includegraphics[width=75mm]{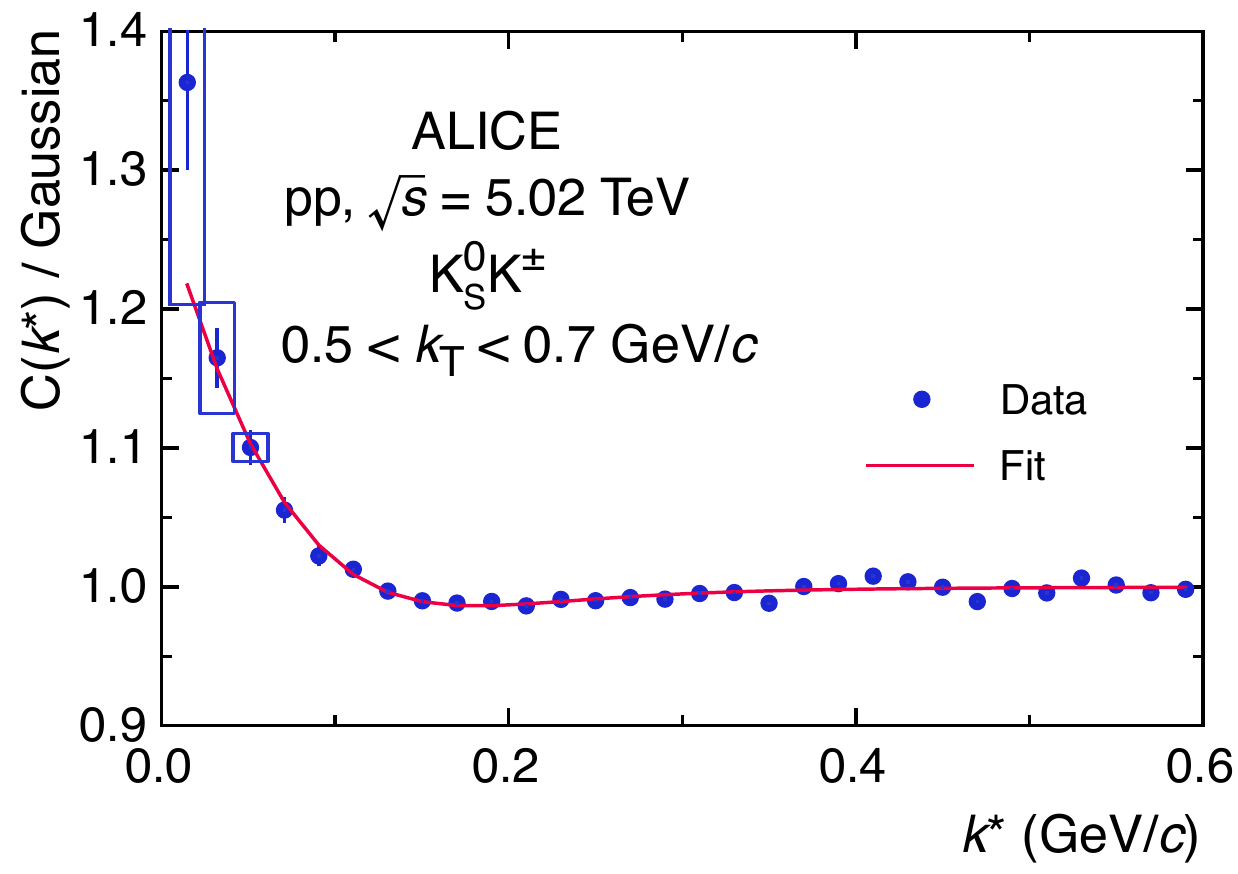}
\hspace{1mm}
\includegraphics[width=75mm]{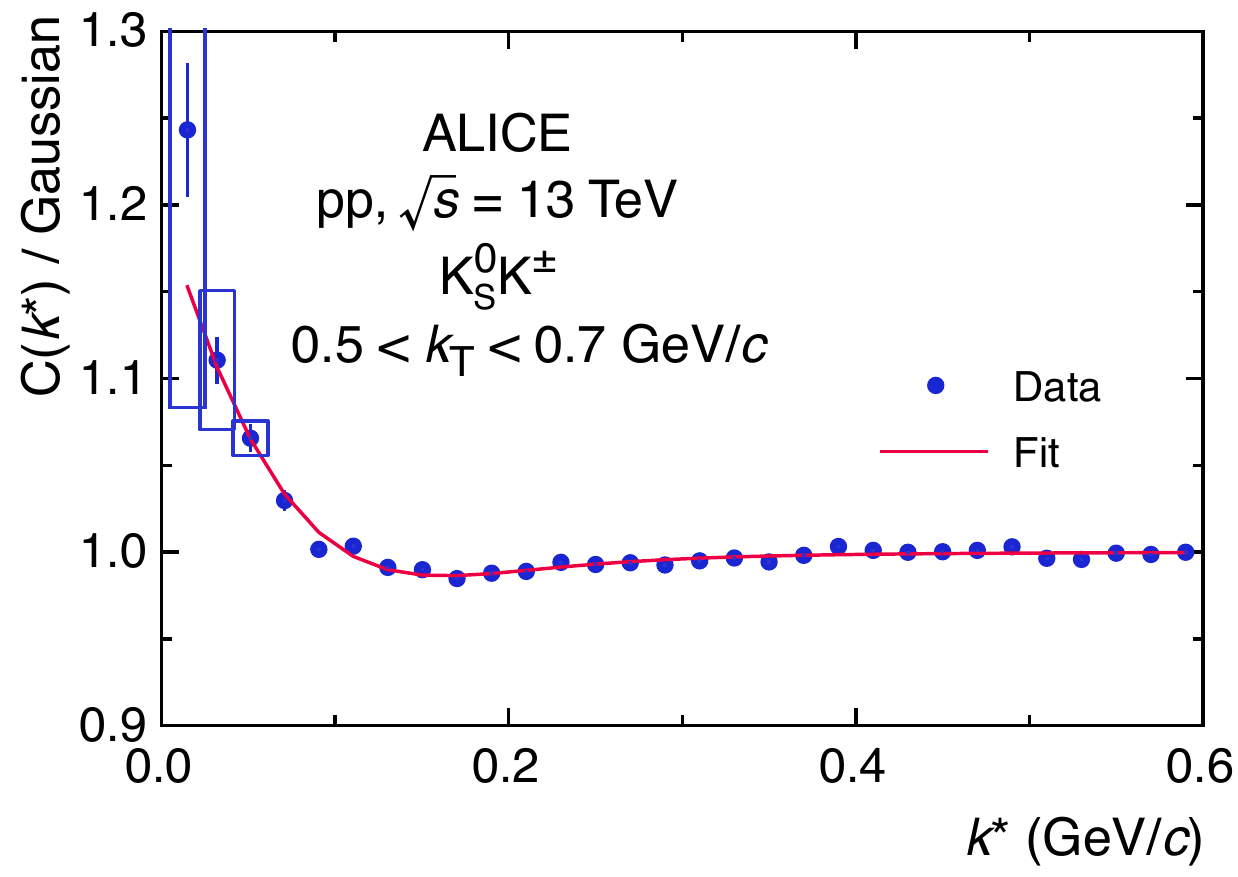}
\caption{Example fits using Eq.~\ref{eq:corrfit} to the ratio of the data in pp collisions at $\sqrt{s}=$
	5.02 TeV (left) and 13 TeV (right) to
	Eq.~\ref{gauss}.
	Statistical uncertainties are shown as error bars and systematic uncertainties are shown as boxes.}
\label{fig:KKchCFfit1}
\end{figure}

Figure~\ref{fig:KKchCFfit1} shows examples of correlation functions divided 
by the Gaussian baseline function, Eq.~\ref{gauss}, with fits of Eq.~\ref{eq:corrfit} for 
K$^0_{\rm S}$K$^{\rm \pm}$, i.e. summed over K$^0_{\rm S}$K$^{\rm +}$ and
K$^0_{\rm S}$K$^{\rm -}$.  
The $a_0$ FSI parameterization coupled with the Gaussian baseline assumption is seen to give a good representation of the signal region of the data, i.e. reproducing the enhancement in the $k^*$ region 0.0--0.1~GeV/$c$ and the small dip in the region 0.1--0.3~GeV/$c$.
The average $\chi^2/$ndf for these fits to the
correlation functions are 1.04 for the left figure and 1.13 for the right figure.
Fits to the data with similarly good $\chi^2/$ndf values are also found using the exponential and quadratic baselines.

\subsection{Systematic uncertainties}
Table~\ref{tab:fitresults} shows the total systematic uncertainties of the extracted $R$ and $\lambda$
parameters from the K$^0_{\rm S}$K$^0_{\rm S}$ and K$^0_{\rm S}$K$^{\rm \pm}$ analyses. The total systematic uncertainty is generally higher than the statistical one. 
The total systematic uncertainty is taken as the square-root of the quadratic sum of the systematic uncertainty
from the fit and the selection criteria.

The fit systematic uncertainty is the combined systematic uncertainty due
to the various baseline assumptions and varying the $k^*$ fit range. 
For K$^0_{\rm S}$K$^0_{\rm S}$, it is calculated from the standard deviation of the
extracted source parameters from six $k^*$ fit ranges: 0.0--0.3, 0.0--0.4, 0.0--0.5, 
0.0--0.6, 0.0--0.7
and 0.0--0.8 GeV/$c$. For K$^0_{\rm S}$K$^{\rm \pm}$ it is calculated from the standard deviation
of using the three baseline functions in four $k^*$ fit ranges: 0.0--0.3, 0.0--0.4, 0.0--0.5 
and 0.0--0.6 GeV/$c$.
The fit values shown in Table~\ref{tab:fitresults} are the average values over these $k^*$ ranges.

The selection systematic uncertainty is the systematic uncertainty related to the
various selection criteria applied in the data analysis.
To determine this, single particle selection criteria were varied by $\sim\pm10$\%, and the value chosen for the minimum separation distance of like charge-sign tracks was varied by 
$\sim 20$\%. The uncertainties in the purity corrections to the $\lambda$ parameters, mentioned
earlier, are also included in the selection systematic uncertainty.
Taking the upper-limit values of the variations to be conservative, this led to additional 
uncertainties of $4\%$ for $R$ and $8\%$ for $\lambda$. As seen in Table~\ref{tab:fitresults}, the fit systematic uncertainty
tends to be comparable to or larger than
the selection systematic uncertainty, reflecting the scale of uncertainties in determining the non-femtoscopic baseline in pp collisions. The ``total quadratic uncertainty'' is the square-root of the
quadratic sum of the ``statistical uncertainty'' column and the ``total systematic uncertainty''
column.

\subsection{Momentum resolution}
\label{subsec:1Dmomres}
Finite track momentum resolution can smear the relative momentum correlation functions used in this analysis. This effect was taken into account using PYTHIA8+GEANT MC simulations. Two PYTHIA8 correlation functions are built using the generator-level momentum ($k^*_{\rm ideal}$) and the measured detector-level momentum ($k^*_{\rm meas}$). Because PYTHIA8 does not incorporate  final-state interactions, weights are calculated using a $9^{th}$-order polynomial fit in $k^*$ to an experimental correlation function and used when filling the same-event distributions. These weights are calculated using $k^*_{\rm ideal}$. Then, the ratio of the ``ideal'' correlation function to the ``measured'' one for each $k^*$ bin is multiplied by the data correlation functions before the fit procedure.
It is found that, due to the large $k^*$ bin size of 20 MeV/$c$ which is used in the analysis of pp collisions,
the correction has a small effect on the lowest $k^*$ bin with the largest statistical error bars, and a negligible effect on the remaining bins. Thus, the momentum resolution 
correction was found to have a $<2\%$ effect on the extracted fit parameters.

\section{Results and discussion}
\begin{figure}[]
\centering
\includegraphics[height=52mm,width=73mm]{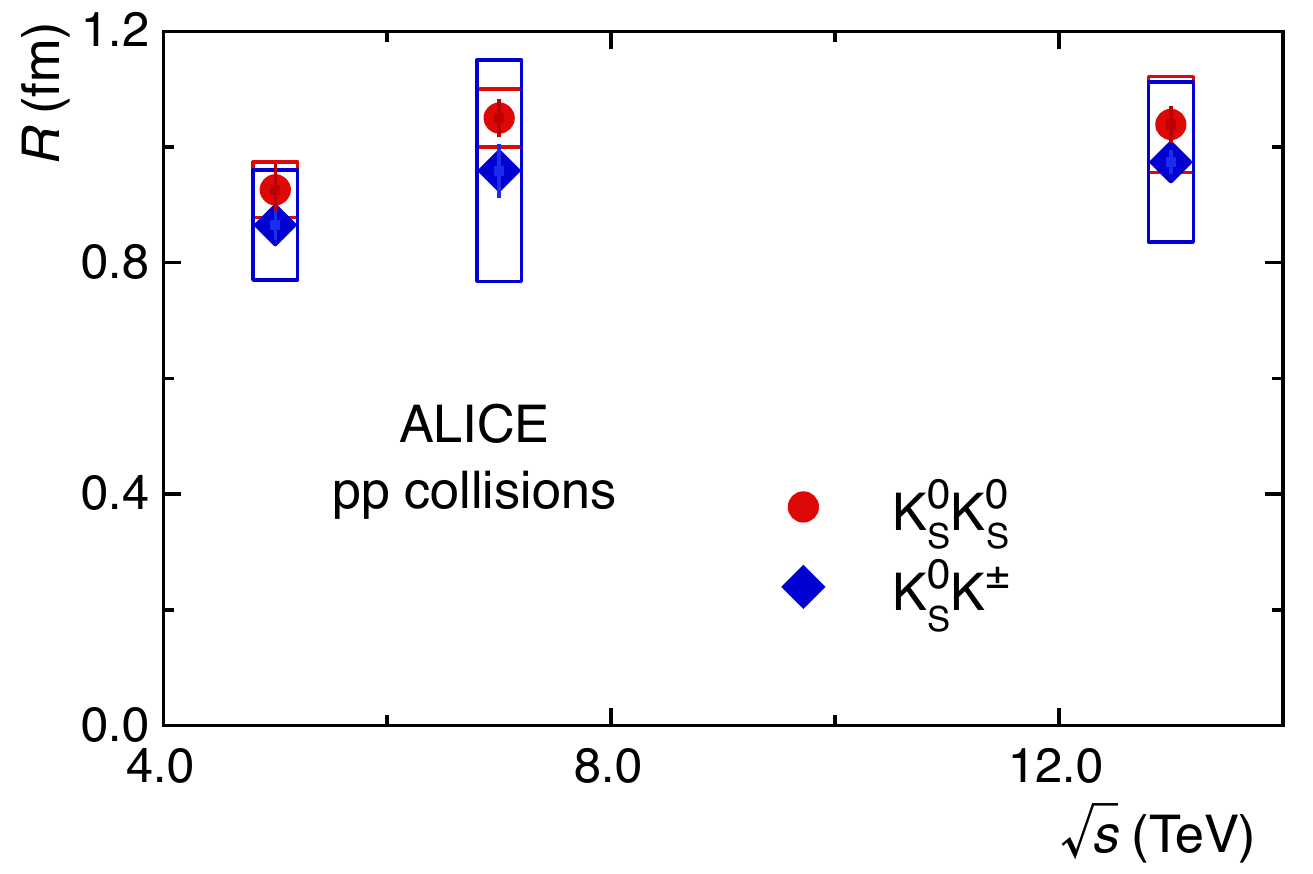}
\hspace{1mm}
\includegraphics[height=50mm,width=75mm]{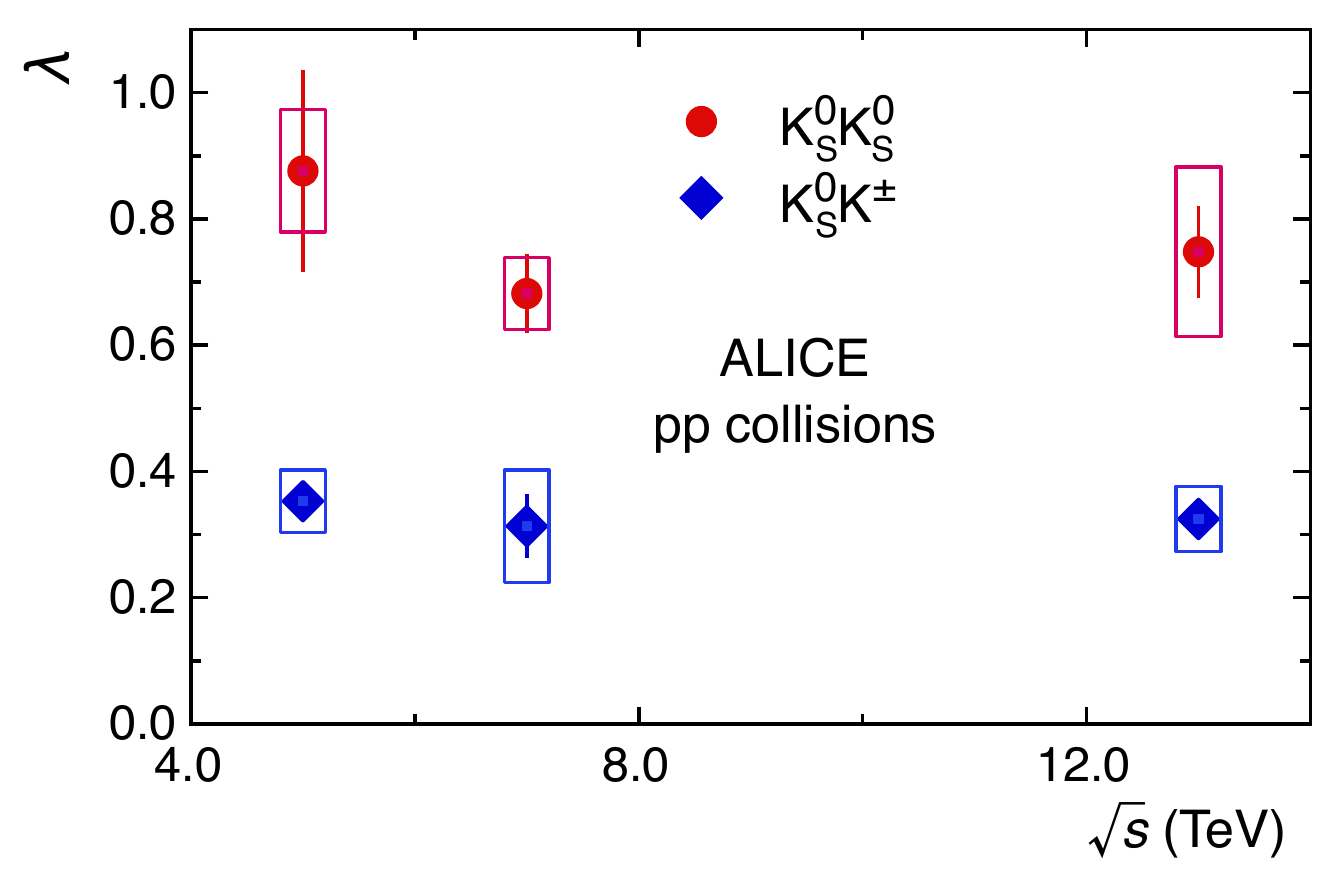}
\caption{$R$ (left) and $\lambda$ (right) parameters extracted in the present analysis 
	from Table~\ref{tab:fitresults}
compared with published K$^0_{\rm S}$K$^0_{\rm S}$ and K$^0_{\rm S}$K$^{\rm \pm}$ results from ALICE 7 TeV pp collisions~\cite{Acharya:2018kpo} averaged over event multiplicity and evaluated 
at $\langle k_{\rm T} \rangle =0.6$ GeV/$c$. Statistical uncertainties
are shown as error bars and
the systematic uncertainties are shown as boxes.}
\label{fig5}
\end{figure}

The extracted source parameters for K$^0_{\rm S}$K$^0_{\rm S}$ 
and K$^0_{\rm S}$K$^{\rm \pm}$, where K$^0_{\rm S}$K$^{\rm +}$ and
K$^0_{\rm S}$K$^{\rm -}$ have been summed over, are shown in Table~\ref{tab:fitresults} and
in Figure~\ref{fig5}. The $\lambda$ parameters are corrected for particle-pair purity.
Figure~\ref{fig5} shows comparisons of the present results for $R$ and $\lambda$
in pp collisions at $\sqrt{s}=$ 5.02 and 13 TeV with
published two-kaon femtoscopic results measured in pp collisions at $\sqrt{s}=$ 7 TeV~\cite{Acharya:2018kpo}.

\begin{table}
 \centering
   \caption{Fit results for average $R$ and $\lambda$ along with statistical and systematic uncertainties. The $\lambda$ parameters are corrected for particle-pair purity.}
 \begin{tabular}{| c || c | c || c | c | c | c || c |}
  \hline
{$R$ or $\lambda$} & pp energy & fit & statistical & fit & selection & total & total \\ 
kaon pair  & (TeV)  & value  & uncert. ($\pm$)  & systematic  & systematic  & systematic  & quadratic  \\ 
    &     &      &     &  uncert. ($\pm$)   & uncert. ($\pm$)  &  uncert. ($\pm$)   & uncert. ($\pm$) \\ \hline\hline
{$R$ (fm)} &  5.02 & 0.926 & 0.045 & 0.031 & 0.037 & 0.048 & 0.066 \\
{K$^0_{\rm S}$K$^0_{\rm S}$}    &  &  &  &  &  &  &  \\ \hline
{$\lambda$} & 5.02 & 0.876 & 0.159 & 0.067 & 0.070 & 0.097 & 0.186 \\
{K$^0_{\rm S}$K$^0_{\rm S}$}    &  &  &  &  &  &  &  \\ \hline\hline
{$R$ (fm)} & 5.02 & 0.865 & 0.025 & 0.088 & 0.037 & 0.095 & 0.098 \\
{K$^0_{\rm S}$K$^{\rm \pm}$}    &  &  &  &  &  &  &  \\ \hline
{$\lambda$} & 5.02 & 0.353 & 0.031 & 0.039 & 0.029 & 0.049 & 0.058 \\
{K$^0_{\rm S}$K$^{\rm \pm}$}    &  &  &  &  &  &  &  \\ \hline\hline
{$R$ (fm)} & 13 & 1.039 & 0.032 & 0.072 & 0.042 & 0.083 & 0.089 \\
{K$^0_{\rm S}$K$^0_{\rm S}$}    &  &  &  &  &  &  &  \\ \hline
{$\lambda$} & 13 & 0.748 & 0.073 & 0.121 & 0.059 & 0.134 & 0.153 \\
{K$^0_{\rm S}$K$^0_{\rm S}$}    &  &  &  &  &  &  &  \\ \hline\hline
{$R$ (fm)} & 13 & 0.974 & 0.020 & 0.131 & 0.042 & 0.138 & 0.139 \\
{K$^0_{\rm S}$K$^{\rm \pm}$}    &  &  &  &  &  &  &  \\ \hline
{$\lambda$} & 13 & 0.325 & 0.020 & 0.044 & 0.028 & 0.052 & 0.055 \\
{K$^0_{\rm S}$K$^{\rm \pm}$}    &  &  &  &  &  &  &  \\ \hline
  \end{tabular}
  \label{tab:fitresults}
\end{table}

For the $R$ parameters, two observations can clearly be made: 
1) there is no significant dependence on $\sqrt{s}$,
i.e. all extracted values are $\sim 1$ fm, and 2) the values extracted from
K$^0_{\rm S}$K$^0_{\rm S}$ and K$^0_{\rm S}$K$^{\rm \pm}$ for a given $\sqrt{s}$ agree
within uncertainties, as would be expected.
It is expected
that $R$ from K$^0_{\rm S}$K$^0_{\rm S}$ and K$^0_{\rm S}$K$^{\rm \pm}$ would agree with each
other if a) the K$^0_{\rm S}$ and K$^{\rm \pm}$ are produced in the same source geometry, 
and b) Eqs.~\ref{eq:fit1} and ~\ref{eq:fit2} properly describe the pair interactions.
Point a) is expected to be true due to isospin invariance of the strong interaction that produces
the kaons in the pp collision, and point b) is supported by the overall good fits that Eqs.~\ref{eq:fit1} 
and ~\ref{eq:fit2} are seen to give to the experimental correlation functions.
The $R$ parameter is essentially independent of $\sqrt{s}$.  While $R$ in general depends on pseudorapidity density also in pp collisions~\cite{Acharya:2019idg,Acharya:2019mzb,Aamodt:2010pp}, the increase expected from the slow logarithmic rise of pseudorapidity density with $\sqrt{s}$ is well within our experimental uncertainties.

\begin{figure}[]
\centering
\includegraphics[height=50mm,width=75mm]{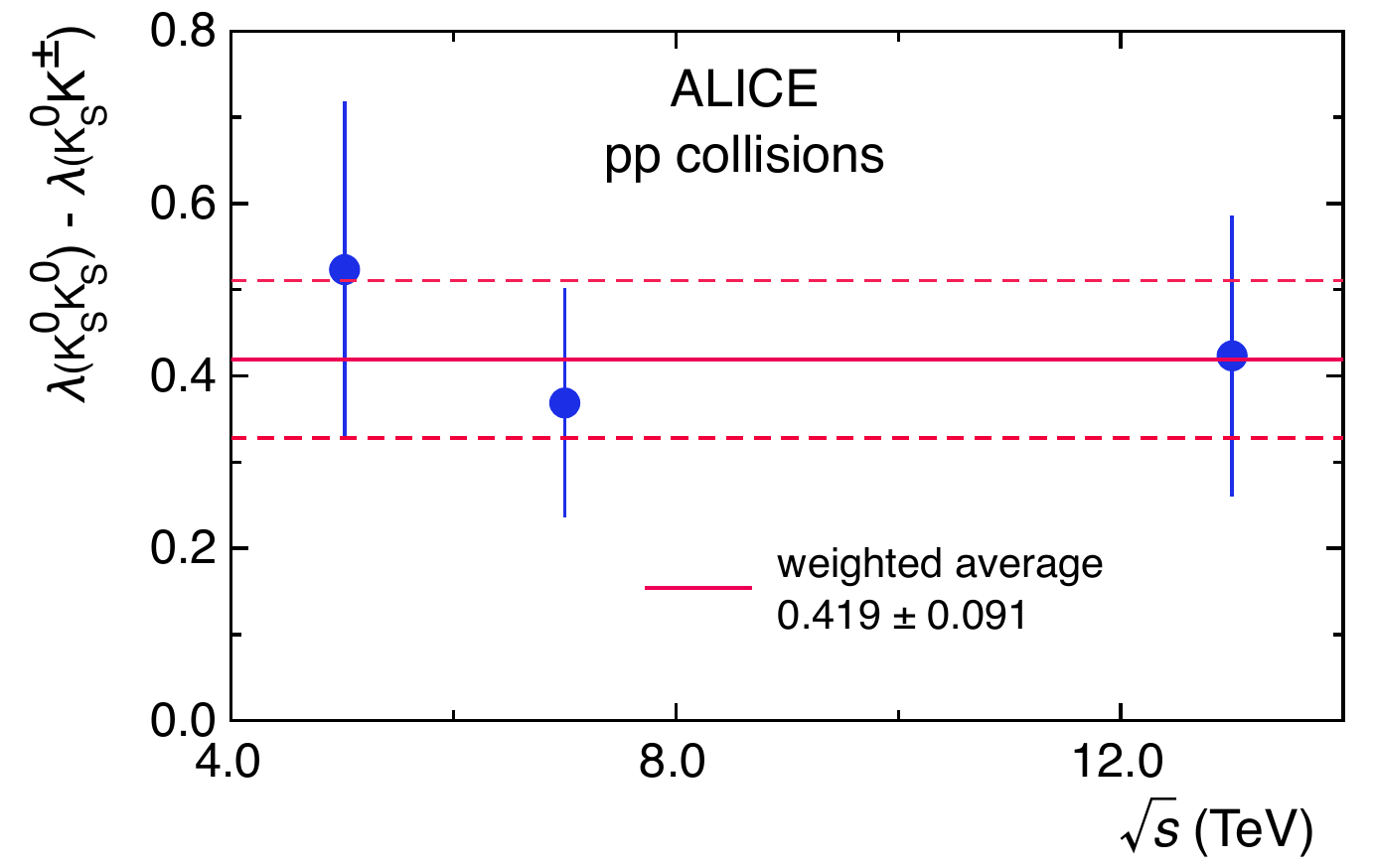}
\caption{Differences in $\lambda$ parameters extracted in the present analysis 
	from Table~\ref{tab:fitresults}
compared with published identical and non-identical kaon results from ALICE 7 TeV pp collisions averaged over event multiplicity and evaluated at $\langle k_T \rangle =0.6$ GeV/$c$. Total
uncertainties are shown. The weighted average of the differences is shown as a solid red line and
the weighted uncertainty, $\pm\sigma$, shown as red dashed lines.}
\label{difflam}
\end{figure}

The extracted $\lambda$ parameters in Fig.~\ref{fig5} suggest that: 1) the values do
not depend significantly on $\sqrt{s}$, 2) the values for K$^0_{\rm S}$K$^0_{\rm S}$ are in the usual range
seen in femtoscopy experiments of $\lambda \sim$ 0.7--0.8, whereas 3) the values for
K$^0_{\rm S}$K$^{\rm \pm}$ are significantly smaller being $\lambda \sim$ 0.3--0.4, consistent with
the 7 TeV results. Figure~\ref{difflam} shows the difference between purity-corrected $\lambda$
parameters extracted with K$^0_{\rm S}$K$^0_{\rm S}$ and K$^0_{\rm S}$K$^{\rm \pm}$
versus $\sqrt{s}$. The propagated total uncertainty is indicated on these points. Also shown is the weighted average of these points, weighted by their total uncertainties. 
It is assumed that the total uncertainties of the K$^0_{\rm S}$K$^0_{\rm S}$ and 
K$^0_{\rm S}$K$^{\rm \pm}$ measurements are uncorrelated. This is considered a reasonable
assumption given the differences in the kaon pairs and the equations used to extract the
source parameters.
The weighted average of the differences
is calculated to be $0.419\pm0.091$, which is $4.6\sigma$ from zero.

There are three main technical factors that, while having a small effect on the $R$ parameter,
can significantly affect the value of the $\lambda$ parameter: 
1) the experimental kaon reconstruction purity,  2) the degree to which a Gaussian distribution describes the kaon source, and 3) the presence of kaons originating from
the decay of long-lived resonances diluting the direct-kaon sample~\cite{Acharya:2018kpo}.
The effect of factor 1) is already corrected for by having divided the extracted $\lambda$ values by
the products of the single-kaon purities given in Section 2.1.
As seen in Figs.~\ref{fig:KKCFfit1} and ~\ref{fig:KKchCFfit1}, the Lednick\'{y} equation, which uses
a Gaussian source, fits the experimental correlation functions well, an observation
that is supported by the good $\chi^2/$ndf values given above, minimizing the effect
of 2). The effects from factor 3) are discussed in the following section.

\subsection{Effect of presence of long-lived resonances}

Table~\ref{tab:resonances} gives a list of mesons from the Review of Particle Physics~\cite{ParticleDataGroup:2020ssz} with masses $< 1500$ MeV/$c^2$ that have decay channels into kaons with significant branching ratios. The two lowest-lying mesons, the K$^*$(892) and the 
$\phi$(1020), are the most abundantly produced and have the narrowest widths, and so are expected to have the greatest effect on the values of the extracted kaon source parameters, which reflect both the kaons produced from the decays of resonances as well as the kaons produced directly from the pp collision. 
Since the mean decay lengths of the K$^*$ and $\phi$ are about $4$ fm and $50$ fm, respectively (see Table~\ref{tab:resonances}), these decays result in effective sources of kaons which are much larger than the expected size of the direct kaon source in pp collisions of about $1$ fm.
Thus, the effect of these should be mostly to
reduce the extracted $\lambda$ parameter. This is due to the correlation function for the smaller-sized direct source being wider in $k^*$ and so
dominating the extracted $R$~\cite{Humanic:2005ye}. Thus, the kaons from these resonances only make an overall suppression of the correlation function.

\begin{table}
 \centering
   \caption{List of mesons from the Review of Particle Physics~\cite{ParticleDataGroup:2020ssz} with 
  masses $< 1500$ MeV/$c^2$ that have decay
  channels into kaons with significant branching ratios.}
 \begin{tabular}{| c | c | c | c | c |}
  \hline
Name & mass (MeV/$c^2$) & $\Gamma$ (MeV/$c^2$) & kaon decays & $\hslash c/ \Gamma$ (fm) \\  \hline
K$^*$(892) &  891.67 & 51.4 & K$\pi$ (100\%) & 3.839 \\ \hline
$\phi$(1020) & 1019.46 & 4.249 & K$^+$K$^-$ (49.2\%), K$^0_{\rm L}$K$^0_{\rm S}$ (34\%) & 46.4 \\ \hline
K$_1$(1270) & 1253 & $\sim100$ & K$\rho$ (42\%) & 1.973 \\ \hline
K$^*_0$(1430) & 1425 & 270 & K$\pi$ ($\sim100$\%) & 0.731 \\ \hline
K$^*_2$(1430) & 1427.3 & 100 & K$\pi$ (50\%), KX (12\%) & 1.973 \\ \hline
  \end{tabular}
  \label{tab:resonances}
\end{table}

The dilution effect on the $\lambda$ parameter due to the K$^*(892)$ and $\phi$(1020) decays can be estimated from K$^{*0}$/K and $\phi$/K ratio measurements from 
ALICE~\cite{Adam:2017zbf,Abelev:2012hy,Acharya:2019bli,Acharya:2019qge}.
Table~\ref{tab:ALICEKphi} shows the measurements of these quantities relevant to the present
estimate. As shown in the table, the measured ratios for both K$^{*0}$/K and $\phi$/K are independent of the collision energy and independent of the decay-kaon charge state within the
measurement uncertainties. For the present calculation, the ratios from 
Refs.~\cite{Acharya:2019bli,Acharya:2019qge} are used since they are taken with an average
$p_{\rm T}$ close to the average $k_{\rm T}$ of 0.6 GeV/$c$ used in this analysis. Whereas the $\phi$
only has one charge state for each of its decay channels, as seen in Table~\ref{tab:resonances}, and 
is its own anti-particle, the K$^*$ has four charge states, and three unique sets of decay charge-state
channels, as shown in Table~\ref{tab:Kstar}, which is taken into account in the present calculation.
Using the numbers in Tables~\ref{tab:resonances}, ~\ref{tab:ALICEKphi}, and ~\ref{tab:Kstar}, the
direct-kaon purity for K$^+$, K$^-$ and K$^0_{\rm S}$, defined as $P$(K$^+$), $P$(K$^-$) and 
$P$(K$^0_{\rm S}$), respectively, where $P$(K$^+$) = $P$(K$^-$) $\equiv P$(K$^\pm$), are calculated
to be $P$(K$^\pm$) $ = 0.726$, and $P$(K$^0_{\rm S})=0.757$.

In the calculations, it has been assumed that the ratios K$^{*0}$/K$^\pm$ = K$^{*\pm}$/K$^\pm$
and K$^{*0}$/K$^0_{\rm S}$ = K$^{*\pm}$/K$^0_{\rm S}$.
The ``diluted'' $\lambda$ parameters can then be estimated for K$^0_{\rm S}$K$^0_{\rm S}$
 as $P$(K$^0_{\rm S}$)$P$(K$^0_{\rm S}$) = $0.57\pm0.02$ and for K$^0_{\rm S}$K$^{\rm \pm}$ as
$P$(K$^0_{\rm S}$)$P$(K$^\pm$) = $0.55\pm0.02$.
The effect of these long-lived resonances is seen to be of the same magnitude, within the uncertainties,
for K$^0_{\rm S}$K$^0_{\rm S}$ and K$^0_{\rm S}$K$^{\rm \pm}$.
The estimate is $\sim$1--2$\sigma$ lower than the purity-corrected 
$\lambda$ values measured
with K$^0_{\rm S}$K$^0_{\rm S}$, however it is $\sim$3--4$\sigma$ larger than the values 
measured in K$^0_{\rm S}$K$^{\rm \pm}$. Thus, the dilution effects on $\lambda$
by the K$^*$(892) and $\phi$(1020) cannot explain the small values for $\lambda$
measured in pp collisions in K$^0_{\rm S}$K$^{\rm \pm}$ femtoscopy.

\begin{table}
 \centering
   \caption{ALICE measurements of K$^*$(892)/K and $\phi$(1020)/K ratios. The uncertainties
   given are the statistical and systematic uncertainties combined in quadrature.}
 \begin{tabular}{| c | c | c | c | c | c |}
  \hline
Ref. & collision & K$^*$(892)/K & $\phi$(1020)/K & average $p_{\rm T}$ (GeV/$c$) & trigger \\  \hline
\cite{Adam:2017zbf} & 2.76 TeV pp & K$^{*0}$/K$^-$ $0.31\pm 0.04$ & $\phi$/K$^-$ $0.11\pm 0.01$ & $\sim 1$ & inelastic \\ \hline
\cite{Abelev:2012hy} & 7 TeV pp & K$^{*0}$/K$^-$ $0.35\pm 0.04$ & $\phi$/K$^-$  $0.11\pm 0.02$ & $\sim 1$ & inelastic \\ \hline
\cite{Acharya:2019bli} & 13 TeV pp & K$^{*0}$/K$^0_{\rm S}$ $0.34\pm 0.01$ & $\phi$/K$^0_{\rm S}$  $0.11\pm 0.01$ &  0.6 & low multiplicity \\ \hline
\cite{Acharya:2019qge} & 5.02 TeV pp & K$^{*0}$/K$^{\pm}$ $0.29\pm 0.02$ & $\phi$/K$^{\pm}$ $0.08\pm0.02$ & $\sim 0.6$ & inelastic \\ \hline
  \end{tabular}
  \label{tab:ALICEKphi}
\end{table}

\begin{table}
 \centering
   \caption{Decay modes of the charge states of the K$^*$(892). Note that the K$^0$ is made up
  of 50\% K$^0_{\rm S}$ and 50\% K$^0_{\rm L}$.}
 \begin{tabular}{| c | c | c |}
  \hline
K$^*(892)$ charge state & decay channels & comment \\  \hline
K$^{*+}$ & K$^+\pi^0$, K$^0\pi^+$ & each channel 50\% \\ \hline
K$^{*-}$ & K$^-\pi^0$, K$^0\pi^-$ & each channel 50\% \\ \hline
K$^{*0}$, $\bar{\rm K}^{*0}$ & K$^+\pi^-$, K$^-\pi^+$, K$^0\pi^0$ & each channel 33.3\% \\ \hline
  \end{tabular}
  \label{tab:Kstar}
\end{table}

\subsection{Physics explanations for differences of $\lambda$ parameters}
Since the technical factors discussed above affecting the extracted $\lambda$ values should affect
the values from K$^0_{\rm S}$K$^0_{\rm S}$ and K$^0_{\rm S}$K$^{\rm \pm}$ in the same way, 
their difference can be ascribed to a physics effect. It is important
to first compare the $\lambda$ parameters extracted in the present work to those measured
in other published KK femtoscopic studies. In Pb--Pb collisions, $\lambda$ is measured to be
$\sim$0.7 for K$^0_{\rm S}$K$^0_{\rm S}$, K$^\pm$K$^\pm$ and 
K$^0_{\rm S}$K$^{\rm \pm}$~\cite{Acharya:2017jks},
similar to what is measured for K$^0_{\rm S}$K$^0_{\rm S}$ presented here, and close to the estimate made for the resonance
dilution effect. For K$^\pm$K$^\pm$ femtoscopy in pp and pPb collisions, $\lambda$ is
measured to be in the range 0.4--0.5, which is smaller than for K$^0_{\rm S}$K$^0_{\rm S}$
presented here~\cite{Acharya:2019fip}. Note that one expects 
K$^\pm$K$^\pm$ to be somewhat smaller than
K$^0_{\rm S}$K$^0_{\rm S}$ on the basis of the resonance dilution effect, since for
K$^\pm$K$^\pm$ the $\lambda$ is estimated to be 
$P$(K$^{\rm \pm}$)$P$(K$^{\rm \pm}$) = 0.53 as compared with
$P$(K$^0_{\rm S}$)$P$(K$^0_{\rm S}$) = 0.57
estimated for K$^0_{\rm S}$K$^0_{\rm S}$. As to why the $\lambda$ parameters in 
K$^\pm$K$^\pm$ in pp and p--Pb collisions are smaller than in Pb--Pb collisions, 
Ref.~\cite{Acharya:2019fip} suggests that this could be due to the kaon source being more
Gaussian in Pb--Pb collisions.

As discussed in
Ref.~\cite{Acharya:2018kpo}, a physics effect that could
cause the difference in $\lambda$ values is related to the 
possibility that the $a_0$ resonance, 
that is solely responsible for
the FSI in the K$^0_{\rm S}$K$^{\rm \pm}$ pair, is a tetraquark state of the form
$(q_1,\overline{q_2}, s, \overline{s})$ instead of a diquark state of the form $(q_1,\overline{q_2})$, 
where $q_1$ and $q_2$ are $u$ or $d$ quarks. The strength of
the FSI through a tetraquark $a_0$ could be decreased by the small source size of the kaon
source, i.e. $R\sim 1$ fm as measured in this analysis, since $s-\overline{s}$ annihilation
would be enhanced due to the close creation proximity. For a FSI through a diquark $a_0$, 
with the form $(q_1,\overline{q_2})$, the small source geometry should not reduce its strength.
For the K$^0_{\rm S}$K$^0_{\rm S}$ case, $\lambda$ would not be affected much by
a tetraquark $a_0$ since the enhancement in the correlation function near $k^*\sim 0$ is
dominated by the effect of quantum statistics.
Note that for the large kaon source measured in Pb--Pb collisions to have $R\sim6$ fm,
the situation would be reversed. The large average separation between the kaons would favor
the formation of a tetraquark $a_0$ and suppress the formation of a diquark $a_0$, and
a larger $\lambda\sim0.6$ is indeed measured in that case, as already mentioned above.
Thus, we can conclude that, as was the case with the published $\sqrt{s}=$ 7 TeV result,  the present results in pp collisions at $\sqrt{s}=$ 5.02 and 13 TeV
are compatible with the $a_0$ being a tetraquark state.

\section{Summary}
In summary, femtoscopic correlations with the particle pair combinations K$^0_{\rm S}$K$^0_{\rm S}$
and K$^0_{\rm S}$K$^\pm$ are studied in pp collisions at $\sqrt{s}=5.02$ and 
$13$ TeV for the first time by the 
ALICE experiment at the LHC. By fitting models
that assume a Gaussian size distribution of the kaon source to the experimental two-particle correlation functions, kaon source parameters are extracted . The model used for the K$^0_{\rm S}$K$^0_{\rm S}$ case includes quantum statistics
and strong final-state interactions through the $f_0$ and $a_0$ resonances. The model used for 
the K$^0_{\rm S}$K$^\pm$ case involves only the final-state interaction through the $a_0$
resonance. In both cases, the models gave a good fit to the experimental correlation functions.
Source parameters extracted in the present work are compared with published
values from ALICE measured in pp collisions at $\sqrt{s}=$ 7 TeV and found to be consistent, i.e. there is no significant dependence
of either $R$ or $\lambda$ on the collision energy.
The new results are compatible with the $a_0$ resonance being a tetraquark state due to the 
$\lambda$ parameter for
K$^0_{\rm S}$K$^\pm$ being significantly smaller than for K$^0_{\rm S}$K$^0_{\rm S}$.


\newenvironment{acknowledgement}{\relax}{\relax}
\begin{acknowledgement}
\section*{Acknowledgements}

The ALICE Collaboration would like to thank all its engineers and technicians for their invaluable contributions to the construction of the experiment and the CERN accelerator teams for the outstanding performance of the LHC complex.
The ALICE Collaboration gratefully acknowledges the resources and support provided by all Grid centres and the Worldwide LHC Computing Grid (WLCG) collaboration.
The ALICE Collaboration acknowledges the following funding agencies for their support in building and running the ALICE detector:
A. I. Alikhanyan National Science Laboratory (Yerevan Physics Institute) Foundation (ANSL), State Committee of Science and World Federation of Scientists (WFS), Armenia;
Austrian Academy of Sciences, Austrian Science Fund (FWF): [M 2467-N36] and Nationalstiftung f\"{u}r Forschung, Technologie und Entwicklung, Austria;
Ministry of Communications and High Technologies, National Nuclear Research Center, Azerbaijan;
Conselho Nacional de Desenvolvimento Cient\'{\i}fico e Tecnol\'{o}gico (CNPq), Financiadora de Estudos e Projetos (Finep), Funda\c{c}\~{a}o de Amparo \`{a} Pesquisa do Estado de S\~{a}o Paulo (FAPESP) and Universidade Federal do Rio Grande do Sul (UFRGS), Brazil;
Ministry of Education of China (MOEC) , Ministry of Science \& Technology of China (MSTC) and National Natural Science Foundation of China (NSFC), China;
Ministry of Science and Education and Croatian Science Foundation, Croatia;
Centro de Aplicaciones Tecnol\'{o}gicas y Desarrollo Nuclear (CEADEN), Cubaenerg\'{\i}a, Cuba;
Ministry of Education, Youth and Sports of the Czech Republic, Czech Republic;
The Danish Council for Independent Research | Natural Sciences, the VILLUM FONDEN and Danish National Research Foundation (DNRF), Denmark;
Helsinki Institute of Physics (HIP), Finland;
Commissariat \`{a} l'Energie Atomique (CEA) and Institut National de Physique Nucl\'{e}aire et de Physique des Particules (IN2P3) and Centre National de la Recherche Scientifique (CNRS), France;
Bundesministerium f\"{u}r Bildung und Forschung (BMBF) and GSI Helmholtzzentrum f\"{u}r Schwerionenforschung GmbH, Germany;
General Secretariat for Research and Technology, Ministry of Education, Research and Religions, Greece;
National Research, Development and Innovation Office, Hungary;
Department of Atomic Energy Government of India (DAE), Department of Science and Technology, Government of India (DST), University Grants Commission, Government of India (UGC) and Council of Scientific and Industrial Research (CSIR), India;
Indonesian Institute of Science, Indonesia;
Istituto Nazionale di Fisica Nucleare (INFN), Italy;
Japanese Ministry of Education, Culture, Sports, Science and Technology (MEXT), Japan Society for the Promotion of Science (JSPS) KAKENHI and Japanese Ministry of Education, Culture, Sports, Science and Technology (MEXT)of Applied Science (IIST), Japan;
Consejo Nacional de Ciencia (CONACYT) y Tecnolog\'{i}a, through Fondo de Cooperaci\'{o}n Internacional en Ciencia y Tecnolog\'{i}a (FONCICYT) and Direcci\'{o}n General de Asuntos del Personal Academico (DGAPA), Mexico;
Nederlandse Organisatie voor Wetenschappelijk Onderzoek (NWO), Netherlands;
The Research Council of Norway, Norway;
Commission on Science and Technology for Sustainable Development in the South (COMSATS), Pakistan;
Pontificia Universidad Cat\'{o}lica del Per\'{u}, Peru;
Ministry of Education and Science, National Science Centre and WUT ID-UB, Poland;
Korea Institute of Science and Technology Information and National Research Foundation of Korea (NRF), Republic of Korea;
Ministry of Education and Scientific Research, Institute of Atomic Physics, Ministry of Research and Innovation and Institute of Atomic Physics and University Politehnica of Bucharest, Romania;
Joint Institute for Nuclear Research (JINR), Ministry of Education and Science of the Russian Federation, National Research Centre Kurchatov Institute, Russian Science Foundation and Russian Foundation for Basic Research, Russia;
Ministry of Education, Science, Research and Sport of the Slovak Republic, Slovakia;
National Research Foundation of South Africa, South Africa;
Swedish Research Council (VR) and Knut \& Alice Wallenberg Foundation (KAW), Sweden;
European Organization for Nuclear Research, Switzerland;
Suranaree University of Technology (SUT), National Science and Technology Development Agency (NSDTA) and Office of the Higher Education Commission under NRU project of Thailand, Thailand;
Turkish Energy, Nuclear and Mineral Research Agency (TENMAK), Turkey;
National Academy of  Sciences of Ukraine, Ukraine;
Science and Technology Facilities Council (STFC), United Kingdom;
National Science Foundation of the United States of America (NSF) and United States Department of Energy, Office of Nuclear Physics (DOE NP), United States of America.
\end{acknowledgement}

\bibliographystyle{utphys}   
\bibliography{KsKs_KsKch.bib}

\newpage
\appendix

%
%

\section{The ALICE Collaboration}
\label{app:collab}
\small
\begin{flushleft}

\bigskip 

S.~Acharya$^{\rm 143}$, 
D.~Adamov\'{a}$^{\rm 97}$, 
A.~Adler$^{\rm 75}$, 
J.~Adolfsson$^{\rm 82}$, 
G.~Aglieri Rinella$^{\rm 34}$, 
M.~Agnello$^{\rm 30}$, 
N.~Agrawal$^{\rm 54}$, 
Z.~Ahammed$^{\rm 143}$, 
S.~Ahmad$^{\rm 16}$, 
S.U.~Ahn$^{\rm 77}$
% ...

\bigskip

\bigskip 

\textbf{\Large Affiliation Notes}

\bigskip 

% ...
$^{\rm VII}$ Also at: University of Kansas, Lawrence, Kansas, United States\\

\bigskip

\bigskip 

\textbf{\Large Collaboration Institutes}

\bigskip 

% ...
$^{16}$ Department of Physics, Aligarh Muslim University, Aligarh, India\\
% ...
$^{30}$ Dipartimento DISAT del Politecnico and Sezione INFN, Turin, Italy\\
% ...
$^{34}$ European Organization for Nuclear Research (CERN), Geneva, Switzerland\\
% ...
$^{54}$ INFN, Sezione di Bologna, Bologna, Italy\\
% ...
$^{75}$ Johann-Wolfgang-Goethe Universit\"{a}t Frankfurt Institut f\"{u}r Informatik, Fachbereich Informatik und Mathematik, Frankfurt, Germany\\
% ...
$^{77}$ Korea Institute of Science and Technology Information, Daejeon, Republic of Korea\\
% ...
$^{82}$ Lund University Department of Physics, Division of Particle Physics, Lund, Sweden\\
% ...
$^{97}$ Nuclear Physics Institute of the Czech Academy of Sciences, \v{R}e\v{z} u Prahy, Czech Republic\\
% ...
$^{143}$ Variable Energy Cyclotron Centre, Homi Bhabha National Institute, Kolkata, India\\
% ...

\bigskip 

\end{flushleft} 
  
\end{document}